\documentclass[a4paper,12pt]{article}
\usepackage[utf8]{inputenc}
\usepackage{overpic}
\usepackage{rotating}
\usepackage{algorithm}
\usepackage{algorithmic}
\usepackage{amsmath}
\usepackage{amssymb}
\usepackage{fancyhdr,lastpage}
\usepackage{enumerate}
\usepackage{graphicx}
\usepackage{epstopdf}
\usepackage[round]{natbib}
\usepackage{rotating}
\usepackage{mathrsfs}
\usepackage{tipa}
\usepackage{hyperref}
\hypersetup{colorlinks,allcolors=black}
\usepackage{authblk}

\newcommand{\bs}[1]{\boldsymbol{#1}}
\newcommand{\pd}[2]{\frac{\partial #1}{\partial #2}}
\newcommand{\td}[2]{\frac{d #1}{d #2}}

\begin{document}
\title{\vspace{-2cm}A thermomechanical model for frost heave and subglacial frozen fringe}
\author[1]{\small Colin R. Meyer}
\author[2]{Christian Schoof}
\author[3]{Alan W. Rempel}
\affil[1]{Thayer School of Engineering, Dartmouth College, Hanover, NH 03755 USA}
\affil[2]{Department of Earth and Ocean Sciences, University of British Columbia, Vancouver, \newline{BC} V6T 1Z4 Canada}
\affil[3]{Department of Earth Sciences, University of Oregon, Eugene, OR 97405 USA}
\maketitle

\begin{abstract}
Ice-infiltrated sediment, known as a frozen fringe, leads to phenomena such as frost heave, ice lenses, and meters of debris-rich ice under glaciers. Understanding the dynamics of frozen fringe development is important as frost heave is responsible for damaging infrastructure at high latitudes and frozen sediments at the base of glaciers can modulate glacier flow, influencing the rate of global sea level rise. Here we study the fluid physics of interstitial freezing water in sediments and focus on the conditions relevant for subglacial environments. We describe the thermomechanics of liquid water flow through and freezing in ice-saturated frozen sediments. The force balance that governs the frozen fringe thickness depends on the weight of the overlying material, the thermomolecular force between ice and sediments across premelted films of liquid, and the water pressure within liquid films that is required by flow according to Darcy's law. We combine this mechanical model with an enthalpy method which conserves energy across phase change interfaces on a fixed computational grid. The force balance and enthalpy model together determine the evolution of the frozen fringe thickness and our simulations predict frost heave rates and ice lens spacing. Our model accounts for premelting at ice-sediment contacts, partial ice saturation of the pore space, water flow through the fringe, the thermodynamics of the ice-water-sediment interface, and vertical force balance. We explicitly account for the formation of ice lenses, regions of pure ice that cleave the fringe at the depth where the interparticle force vanishes. Our model results allow us to predict the thickness of a frozen fringe and the spacing of ice lenses at the base of glaciers.
\end{abstract}

\section{Introduction}
Freezing of interstitial water in sediments commonly occurs in subaerial and subglacial environments, contributing to effects such as frost heave, needle ice, and the transport of subglacial debris \citep{Hem2004,Das2006,Wet2006}. Through multiple cycles of freeze and thaw in high latitude environments, patterns can develop such as the arctic stone circles \citep{Kes2003}. In this paper, we consider the thermodynamical and fluid dynamical processes that occur as water freezes in a porous medium. We describe the melting and freezing processes using an enthalpy formulation, which facilitates our numerical method as we avoid tracking phase change interfaces and elucidates the role of the frozen fringe as a mushy zone between ice- and water-saturated sediments. Our treatment is general enough to apply in a variety of industrial and environmental contexts where interstitial freezing occurs, yet here we primarily focus on geophysical applications. 

Consider frost heave, the common freeze/thaw phenomenon that takes place throughout high latitudes. As water held within sediments freezes, ice lenses cleave the sediment and expand, causing vertical displacement of the ground surface. Such surface displacement causes significant damage to infrastructure at high latitudes and is due to the growth of distinct ice lenses within the soil rather than the water density change on freezing. \citet{Tab1930} demonstrated this key fact by freezing a sediment pack saturated with benzene, which contracts on freezing; the benzene produced significant heave through the expansion of discrete ice lenses. 

Early models for frost heave relied on surface tension to draw water to the lowest ice lens, i.e. the so-called ``primary model for frost heave''. This model suffers from several deficiencies, most importantly that surface tension acts tangential to the ice surface and cannot provide the upward force to drive heave. In addition, there is no mechanism to form distinct ice lenses in primary heave, which led \citet{O'Ne1985} to derive the ``secondary model of frost heave,'' wherein a zone of partially ice saturated sediment extends below the lowest ice lens. \citet{Fow1994} clarified the mathematical model for secondary frost heave and analysed an asymptotically reduced form of the model. \citet{Rem2004} highlighted the role of premelting at the interface between ice and sediment grains. The disjoining pressure across the liquid between ice and sediment grains relates the local melting temperature to the vertical force balance. \citet{Fow1994}, on the other hand, choose the liquid pressure to be given as a function of soil water content as set by surface tension, reminiscent of a primary frost heave model. In what follows, we build on the \citet{Rem2004} formulation, highlighting an alternate derivation, writing out the equations, and systematically reducing the equations asymptotically, similar to \citet{Fow1994}. 

Field observations show several meters of frozen sediment are commonly attached to the base of glaciers, motivating efforts to understand glaciohydraulic supercooling \citep[e.g.][]{Rot1987,Law1998,Cre2013} and frost heave. In the fastest flowing reaches of glaciers and ice sheets, sliding dominates glacier motion and the rate of sliding is tied to the temperature and water pressure at the glacier base --- the same things that control the growth of a frozen fringe. \citet{Rem2008} treats the overlying glacier as a large lowest ice lens and predicts meters-scale frozen fringes below glaciers for typical parameters, in line with observations. 

\citet{And2012,And2014} analyse freezing colloid suspensions using a directional solidification experiments with aqueous suspensions of alumina particles. \citet{And2014} developed a model built on the \citet{Rem2004} framework that include compaction and cohesion of the colloid suspension. The \citet{And2014} model assumes a steady state linear temperature profile throughout the experiment and boils down to a system of ordinary differential equations for the location of the compaction front and frozen fringe extent, reminiscent of \cite{Fow1993}. Based on their model, \citet{And2014} describe a regime diagram showing the three primary freezing regimes observed in their experiments: periodic ice lenses, disordered ice lenses, and periodic ice banding.

Frozen fringes and freeze-thaw cycles are inherently problems of phase change and partial melting. In these types of problems, it can be valuable to solve the energy conservation equation in an enthalpy form rather than for the temperature to avoid explicitly tracking phase change interfaces. The enthalpy (i.e. sum of the sensible and latent heat) accommodates the phase change, which facilitates numerical solutions. From sea ice \citep{Kat2008b}, permafrost \citep{Clo2018} and meltwater percolation through snow \citep{Mey2017c} to industrial processes \citep{Vol1987}, the enthalpy approach to phase change problems is useful for many applications. Enthalpy methods have been used extensively for polythermal glaciers, where part of the glacier is below the melting point and the rest of the glacier is at the melting point, i.e. temperate ice \citep{Asc2012,Sch2016}. Here we use the enthalpy method to solve for energy conservation within a frozen fringe. 

In this paper, we focus on the conditions relevant for subaerial frost heave and subglacial environments. Although numerous treatments of frost heave exist in the literature \citep[e.g.][]{O'Ne1985,Fow1994,Rem2004}, here we derive our model from scratch for completeness and clarity. In section \ref{sec:model}, we start by writing down mass, momentum, and energy conservation equations for a frozen fringe. Then, we nondimensionalise and systematically reduce the equations by exploiting the small density difference between ice and water as well as the large latent heat of fusion upon freezing water (i.e. a large Stefan number). We solve our reduced model using an enthalpy method, where phase-change boundaries are determined implicitly on a fixed grid. In section \ref{sec:results}, we demonstrate the results of our enthalpy model. We analyse a steady state frozen fringe thickness in melting and balanced thermodynamic conditions in both a semi-analytical model and an enthalpy framework. Then, we examine the local effective pressure for melting and freezing conditions, highlighting ice lens formation. Lastly, we show the formation of periodic ice lenses and map out the different behaviour in a regime diagram for the heave rate and effective pressure. Finally, we offer conclusions and discuss future directions in section \ref{sec:conclusions}. 

\begin{figure}
    \centering
    \includegraphics[width=\linewidth]{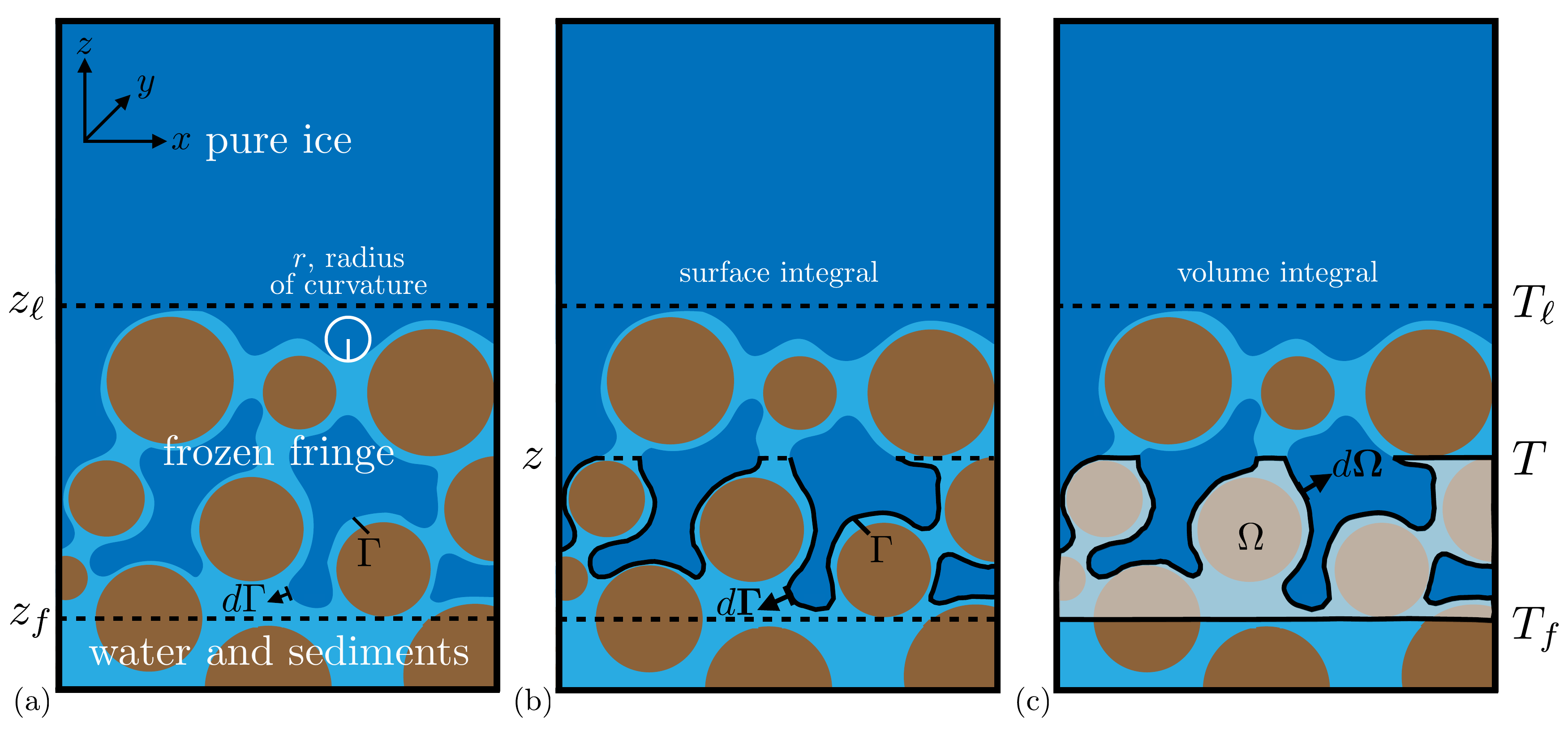}
    \caption{Schematic of the a frozen fringe system: (a) components including pure ice (lowermost ice lens or the bottom of a glacier), frozen fringe, and the unfrozen porous mixture of water-saturated sediments \citep[after][]{Rem2004,And2014}. (b) surface integral path $\Gamma$ and inward-pointing normal vector $d\boldsymbol{\Gamma}$. (c) volume integral domain $\Omega$ with the outward-pointing normal vector $d\boldsymbol{\Omega}$.}
    \label{fig:frozenfringeschematic}
\end{figure}

\section{Model\label{sec:model}}
Inside the frozen fringe, which is shown schematically in figure \ref{fig:frozenfringeschematic}, we define a coordinate system that is fixed with respect to the immobile, water-saturated sediment below, with $z$ vertical, $x$ the lateral coordinate, and $y$ pointing into the page. We label the deepest extent of the fringe as $z=z_f$ and the top of the fringe as $z=z_{\ell}$, or equivalently $z_{\ell} = z_f+h$, where $h$ is the fringe thickness. 

\subsection{Mass conservation}
The frozen fringe is partitioned into three components: ice, water, and sediment. The porosity $\phi$ denotes the volume of voids (i.e. ice and water) within a representative control volume. The fraction of the voids that is taken up by ice is the ice saturation $S$. Mass conservation for sediment, ice, and water implies that
\begin{eqnarray}
\pd{\left[\rho_s(1-\phi)\right]}{t} + \boldsymbol{\nabla}\cdot \left[ \rho_s (1-\phi) \boldsymbol{V_\textrm{s}}\right] &=& 0,~~~(\mbox{sediment})\label{sedcons}\\
\pd{\left(\rho_i\phi S\right)}{t} + \boldsymbol{\nabla}\cdot \left[ \rho_i\phi S \boldsymbol{V}\right] &=& -m,~~~(\mbox{ice})\label{eqn:icecons}\\
\pd{\left[\rho_w\phi (1-S)\right]}{t} + \boldsymbol{\nabla}\cdot \left[ \rho_w\phi \left(1-S\right) \boldsymbol{U}\right] &=& m,~~~~~(\mbox{water})\label{eqn:watercons}
\end{eqnarray}
where $m$ is the rate at which ice (density $\rho_i$) is melted, i.e. converted into liquid water (density $\rho_w$), $\boldsymbol{V}$ is the speed at which the ice moves through the fringe due to heaving at the ice lens above the fringe, and $\boldsymbol{V_\textrm{s}}$ is the sediment (density $\rho_s$) velocity. The water flux through the fringe is $\boldsymbol{U}$, which is given by Darcy's law as
\begin{equation}
\phi \left(1-S\right) \boldsymbol{U} = -\frac{k}{\mu} \left[\boldsymbol{\nabla}p_w+\rho_w g \boldsymbol{\hat{z}}\right],
\label{eqn:darcyfirst}
\end{equation}
where the permeability $k$ depends on the ice saturation as well as the porosity and other properties of the sediment matrix. Here $g$ is the acceleration due to gravity. Values for all of the parameters are given in table \ref{prms}.

In this paper, we assume that the porosity is constant throughout the fringe. Both below the frozen fringe and in its interior, there may be significant compaction due to the reduced water pressure as the lens pulls in water to freeze \citep{Fow1999, And2012, And2014}. We neglect such complications for now and take the entire sediment pack to maintain a constant porosity $\phi$ that jumps to $\phi=1$ at ice lenses. A new ice lens forms when the force between sediment grains reaches zero, as described in the next section. 

\subsection{Force balance}
The force balance within the fringe is composed of three components: the weight of the material above and within the fringe, the water pressure within and below the fringe, as well as the thermomolecular force between sediment grains and interstitial ice, which acts across a thin film of premelted water \citep{Das2006}. 
Integrating these components over the surface area of the fringe gives
\begin{equation}
\int_{\Gamma}{p_i \boldsymbol{n}~d\Gamma} = \int_{\Gamma}{\left(p_i - p_w\right) \boldsymbol{n}~d\Gamma} +\int_{\Gamma}{p_w \boldsymbol{n}~d\Gamma},\label{forcebalint}
\end{equation}
in which $p_i$ is the isotropic ice pressure (i.e. local normal stress) at the ice-water interface within the fringe, $p_w$ is the water pressure at the boundary, and $\boldsymbol{n}$ is the inward-pointing unit normal to the boundary $\Gamma$ (cf. figure \ref{fig:frozenfringeschematic}). The difference between the ice and water pressure, i.e. the first term on the right hand side of equation \eqref{forcebalint}, is accommodated by a thermomolecular force \citep{Rem2001,Rem2004,Wet2006}.

We convert these surface integrals over $\Gamma$ to volume integrals over the unfrozen component of the fringe $\Omega$. We construct a closed surface by adding surface integrals with flat surfaces at $z_f$ and $z$, as shown schematically in figure \ref{fig:frozenfringeschematic}. That is, for a generic pressure field $p$, we have the integrals 
\begin{equation}
\int_{\Omega}{\nabla p~d\Omega} = \int_{\Gamma}{p \boldsymbol{n}~d\Gamma} + \int_{\Gamma_z}{p \boldsymbol{n}~d\Gamma} + \int_{\Gamma_{z_f}}{p \boldsymbol{n}~d\Gamma},\label{genericintcap}
\end{equation}
where $\boldsymbol{n}$ is the outward-pointing normal for the volume $\Omega$. In other words, $\boldsymbol{n}=\boldsymbol{k}$ on the upper cap at $z$, the lowest ice lens, and $\boldsymbol{n}=-\boldsymbol{k}$ on the lower cap at the bottom of the fringe $z_{f}$, where $\boldsymbol{k}$ is the unit vector in the $z$-direction.

We take the surface $\Gamma_z$ to be the ice boundary at some height $z$, which we can write $\left|\Gamma_z\right| = (1-\phi S)A$ where $A$ is the cross-sectional area. This surface has the two limits of $\left|\Gamma_{z_{\ell}}\right| = 0$ at the bottom boundary of the lowest active ice lens ($\phi=1$, $S=1$) and $\left|\Gamma_{z_{f}}\right| = A$ at the bottom of the fringe ($S=0$). For this reason, no upper cap is necessary when integrating across the entire fringe, and equation \eqref{genericintcap} reduces to 
\begin{equation}
\int_{\Omega}{\nabla p~d\Omega} = \int_{\Gamma}{p \boldsymbol{n}~d\Gamma} + \int_{\Gamma_{z_f}}{p \boldsymbol{n}~d\Gamma},\label{genericintcapfull}
\end{equation}

We assume that water flow through ice-saturated porous fringe is governed by Darcy's law and that the water pressure varies only on a lengthscale set by the fringe and not on the scale of individual grains. This assumption allows us to define the water pressure throughout the volume $\Omega$, even though part of the domain is filled with ice and sediment. Crucially, we follow \citet{Rem2004} and assume that the microscale pressure is the homogenised Darcy pressure. In other words, we treat the water pressure in the thin films between sediment and ice as well as the water pressure in the pore throats between sediment grains as determined by Darcy's law, which is a key difference between \citet{Fow1994} and \citet{Rem2004}.

At this stage, we restrict our focus to a one-dimensional water pressure that only depends on the vertical coordinate $z$ and assume that there are no transverse pressure gradients. Therefore, inserting the water pressure $p_w$ into equation \eqref{genericintcap}, we find that
\begin{equation} 
\int_{\Gamma}{p_w \boldsymbol{n}~d\Gamma}  = A\left[ \int_{z_f}^{z}{(1-\phi S)\pd{p_w}{z'}dz'} + p_w(z_f) - (1-\phi S)p_w(z)\right]\boldsymbol{k},\label{waterpressure}
\end{equation}
where the porosity $\phi$ and saturation $S$ can also depend on the vertical coordinate $z$. Now integrating across the entire fringe from $z=z_f$ to $z=z_{\ell}$ gives
\begin{equation} 
\int_{\Gamma}{p_w \boldsymbol{n}~d\Gamma}  = A\left[ \int_{z_f}^{z_{\ell}}{(1-\phi S)\pd{p_w}{z'}dz'} + p_w(z_f)\right]\boldsymbol{k}.\label{waterpressurefull}
\end{equation}
The analogous equation for  $p_i-p_w$ represents the thermomolecular contribution to the force balance \citep{Rem1999,Rem2004} and is given by 
\begin{equation} 
\int_{\Gamma}{\left(p_i-p_w\right) \boldsymbol{n}~d\Gamma}  = A\left[ \int_{z_f}^{z_{\ell}}{(1-\phi S)\pd{\left(p_i - p_w\right)}{z'}dz'} + p_i(z_f)-p_w(z_f)\right]\boldsymbol{k}.\label{eqn:thermomolecular}
\end{equation}

When integrated across the entire fringe, the water pressure and thermomolecular force balance the total normal stress $\sigma_n$ at $z_f$ resulting from the weight of the overlying material (e.g. equation \eqref{forcebalint} and figure \ref{fig:frozenfringeschematic}). With this in mind, we write
\begin{equation} 
\int_{\Gamma}{p_i \boldsymbol{n}~d\Gamma} = \left\{\sigma_n - \int_{z_f}^{z_{\ell}}{\left[\rho_s g (1-\phi) + \rho_w g \phi (1-S)\right]dz'} \right\} A \boldsymbol{k},\label{overburden}
\end{equation}
where we have included the weight of fringe material as the integral over each constituent.

We now combine all of these equations including the effects of gravity and arrive at 
\begin{eqnarray} 
\sigma_n = \int_{z_f}^{z_{\ell}}{\left[\rho_s g (1-\phi) + \rho_w g \phi (1-S)\right]dz'} + \hspace{7.5cm}\nonumber \\ 
\int_{z_f}^{z_{\ell}}{(1-\phi S)\pd{\left(p_i-p_w\right)}{z'}dz'} + p_i(z_f) -p_w(z_f) + \hspace{2cm}\nonumber \\ \int_{z_f}^{z_{\ell}}{(1-\phi S)\pd{p_w}{z'}dz'} + p_w(z_f).\hspace{0cm}\label{fullforce}
\end{eqnarray}
We define the effective pressure at the base of the fringe $N$ as the total normal stress $\sigma_n$ at $z_f$ supported by the fringe less the water pressure at the base of the fringe $p_w(z_f)$ so that \begin{eqnarray} 
N = \int_{z_f}^{z_{\ell}}{\left[\rho_s g (1-\phi) + \rho_w g \phi (1-S)\right]dz'} + \hspace{7.5cm}\nonumber \\ 
\int_{z_f}^{z_{\ell}}{(1-\phi S)\pd{\left(p_i-p_w\right)}{z'}dz'} + p_i(z_f) -p_w(z_f) + \int_{z_f}^{z_{\ell}}{(1-\phi S)\pd{p_w}{z'}dz'}.\label{esfullforce}
\end{eqnarray}
We recognise $N$ as the load supported by contacts between sediment grains at $z_f$.

Additionally, we define the local effective pressure $N_{\textrm{loc}}(z)$ as the portion of the overlying load that is supported at a height $z$ by sediment grain contacts. The rest of the overlying load is supported by thermomolecular forces or water pressure acting on the ice fringe below the height $z$ as well as water pressure at height $z$. The thermomolecular and water pressure contributions from below $z$ are given as 
 \begin{eqnarray} 
\int_{\Gamma}{p_i \boldsymbol{n}~d\Gamma}  = A\left\{\int_{z_f}^{z}{\left[\rho_s g (1-\phi) + \rho_w g \phi (1-S) \right]dz'} \hspace{6.25cm}\right. \nonumber \\  \left.+ \int_{z_f}^{z}{(1-\phi S)\pd{(p_i-p_w)}{z'}dz'} + p_i(z_f)-p_w(z_f) - (1-\phi S)\left[ p_i(z)-p_w(z) \right] \right. \hspace{0.75cm}\nonumber \\ 
\left. \int_{z_f}^{z}{(1-\phi S)\pd{p_w}{z'}dz'} + p_w(z_f)  - (1-\phi S)p_w(z) \right\}\boldsymbol{k}.\label{below}
\end{eqnarray}
We assume that sediment grains have infinitesimal contacts and, therefore, the water pressure at the height $z$ supports the force $A\left(1-\phi S\right)p_w(z)\boldsymbol{k}$, which excludes the areas occupied by ice. The total force supported by grain contacts at a height $z$ is the overburden $\sigma_n$ minus both equation \eqref{below} and the water pressure at $z$. Thus, we can write the effective pressure $N_{\textrm{loc}}(z)$ as  
\begin{eqnarray} 
N_{\textrm{loc}}(z) = N - \left\{\int_{z_f}^{z}{\left[\rho_s g (1-\phi) + \rho_w g \phi (1-S)\right]dz'} \hspace{6cm}\right. \nonumber \\  \left.-\int_{z_f}^{z}{\phi S\pd{(p_i-p_w)}{z'}dz'} + \phi S\left[ p_i(z)-p_w(z) \right] + \int_{z_f}^{z}{(1-\phi S)\pd{p_w}{z'}dz'}  \right\}.\label{localN}
\end{eqnarray} 
A new ice lens initiates at the height $z_n$ where the local effective pressure is zero, i.e. $N_{\textrm{loc}}(z_n) = 0$, as there is no longer any force on the sediment grains \citep{O'Ne1985,Rem2004,And2014}. We treat the effective pressure at the bottom of the fringe $N$ as an input to the model that is determined by groundwater hydrology or subglacial drainage \citep[e.g.][]{Sch2010b}.

\subsection{Generalized Clausius-Clapeyron and Gibbs-Thomson}
The pressure difference between ice and water is related to temperature through the generalised Clausius-Clapeyron equation, which in its linearised form is given by
\begin{equation}
p_i - p_w = \rho_i \mathscr{L} \frac{T_m-T}{T_m} + \left( p_m - p_w\right) \frac{\rho_w-\rho_i}{\rho_w},
\end{equation}
where the bulk melting temperature at the reference pressure $p_m$ is $T_m$, the specific latent heat of fusion for ice is $\mathscr{L}$, and the densities of ice and water are given as $\rho_i$ and $\rho_w$, respectively \citep{Wor1999, Wor2000, Cla2005}. We choose the reference pressure to be the overburden $\sigma_n$, so that at the bottom of the fringe $z=z_f$, we have
\begin{equation}
p_i(z_f) - p_w(z_f) = \rho_i \mathscr{L} \frac{T_m-T}{T_m} + \frac{\rho_w-\rho_i}{\rho_w}N,
\end{equation}
and in the interior of the fringe, we have
\begin{equation}
p_i(z) - p_w(z) = \rho_i \mathscr{L} \frac{T_m-T}{T_m} + \frac{\rho_w-\rho_i}{\rho_w}\left[N + p_w(z_f) - p_w(z)\right].\label{eqn:clausius}
\end{equation}
Note that in this formulation changes to $\sigma_n$ affect the value of $N$ and $T_m$. 

At the bottom of the fringe, the ice is in contact with water in between pore throats, as shown in the schematic in figure \ref{fig:frozenfringeschematic}. The curvature induced by the space between sediment grains, leads to a difference in the pressure in ice and water phases due to the Gibbs-Thomson effect \citep{Wor2000}, which is given by 
\begin{equation}
p_i(z_f) - p_w(z_f) = \gamma \kappa,
\label{eqn:GibbsThomson}
\end{equation}
where $\gamma$ is the ice-water surface energy and $\kappa$ is the curvature of the ice-water interface. Moreover, the curvature $\kappa$ at the bottom of the fringe is related to the radius of curvature for sediment pore throats $r_p$ as $\kappa=2/r_p$. The critical effective pressure required to overcome the pore throat curvature is given by the Gibbs-Thomson effect and defined by the right-hand side of equation \eqref{eqn:GibbsThomson}, i.e.
\begin{equation}
N_c = \frac{2\gamma}{r_p},
\end{equation}
\citep[e.g.][]{Fow1997,Rem2008,Mey2018c}.

Combining the generalised Claussius-Clapeyron equation \eqref{eqn:clausius} and the Gibbs-Thomson effect \eqref{eqn:GibbsThomson} at the bottom of the fringe, we can relate the curvature induced by pore throats to the temperature at the interface, which is given by
\begin{equation}
\rho_i \mathscr{L} \frac{T_m-T}{T_m} = N_c - \frac{\rho_w-\rho_i}{\rho_w}N.
\label{eqn:CCandGT}
\end{equation}
The contribution from surface energy on the right hand side typically dominates the term proportional to the density difference, since ice and water densities differ by less than 10\%. Therefore, it is useful to define the undercooling temperature $T_f$ that supports this balance as
\begin{equation}
T_f = T_m-\frac{N_c T_m}{\rho_i \mathscr{L}},
\end{equation}
which allows us to write equation \eqref{eqn:CCandGT} as
\begin{equation}
T(z_f) = T_f + \frac{\left(\rho_w-\rho_i\right)N}{\rho_i \rho_w \mathscr{L}}T_m.
\end{equation}
\citet{Rem2008} drops the second term on the right arguing that it is small, which is consistent with our dominant balance above. We, however, keep all terms for now and reduce the model systematically in \S\ref{reduction}.

At this stage, we can now insert the generalised Clausius-Clapeyron equation \eqref{eqn:clausius} and the Gibbs-Thomson effect \eqref{eqn:GibbsThomson} into the effective pressure integrals \eqref{esfullforce} and \eqref{localN}. The effective pressure at the bottom of the fringe $N$ is then
\begin{eqnarray} 
N = N_c+\int_{z_f}^{z_{\ell}}{\left[\rho_s g (1-\phi) + \rho_w g \phi (1-S) \right]dz'} - 
\int_{z_f}^{z_{\ell}}{(1-\phi S)\left[\frac{\rho_i \mathscr{L}}{T_m}\pd{T}{z'}-\frac{\rho_i}{\rho_w}\pd{p_w}{z'}\right]dz'}.\label{esfullforce_T}
\end{eqnarray}
In the interior of the fringe, the local effective pressure is given by
\begin{eqnarray} 
N_{\textrm{loc}}(z) = N - \left\{\int_{z_f}^{z}{\left[\rho_s g (1-\phi) + \rho_w g \phi (1-S)\right]dz'} +\frac{\rho_i \mathscr{L}}{T_m} \int_{z_f}^{z}{\phi S \pd{T}{z'} dz'} + \right. \hspace{2cm}\nonumber  \\ \left. \phi S\left[ \rho_i \mathscr{L} \frac{T_m-T}{T_m} + \frac{\rho_w-\rho_i}{\rho_w}\left[N + p_w(z_f) - p_w(z)\right] \right] + \int_{z_f}^{z}{\left(1-\frac{\rho_i}{\rho_w} \phi S \right)\pd{p_w}{z'}dz'}  \right\}.\label{localN_T}
\end{eqnarray}
which connects the local pressure and temperature. 

If there is not a fringe, the development of the force balance at the base of the fringe still holds, except that $z_{f} = z_{\ell}$ and the integrals vanish. Also, the curvature at the bottom of the ice is no longer $\kappa = 2/r_p$ and so the effective pressure is given by
\begin{equation}
N = p_i(z_f) - p_w(z_f) < N_c, 
\end{equation}
while the temperature at bottom of the ice is given by 
\begin{equation}
 T(z_f) = T_m - \frac{T_m}{\rho_w \mathscr{L}}N,
\end{equation}
which is modulated by the effective pressure. 

\subsection{Energy conservation}
Mass exchange between the liquid and solid phases within the fringe leads to changes in ice saturation. Conservation of energy determines the temperature and phase change within the fringe and can be expressed in terms of specific enthalpy, $h_\alpha$. For the constituents $\alpha$ that make up the fringe, sediment ($s$), ice ($i$), and water ($w$), conservation of energy in enthalpy form is given as
\begin{eqnarray}
\int_{\Omega}{\pd{}{t} \left[ (1-\phi)\rho_s h_s + \phi S\rho_{i}h_{i} + \phi (1-S) \rho_{w}h_{w}\right] d\Omega} =\hspace{6cm} \nonumber \\ \int_{\Gamma}{ \left\{ -\phi S \rho_i h_i \boldsymbol{V} - \phi \left( 1- S\right) \rho_{w}h_{w}\boldsymbol{U} - \left( 1-  \phi \right)\rho_{s}h_{s}\boldsymbol{V}_s  + K_e\boldsymbol{\nabla}T\right\}\cdot d\boldsymbol{\Gamma}},\label{eqn:enthalpy1}
\end{eqnarray}
where $K_e$ is the effective thermal conductivity \citep[Appendix B of][]{Rem2008}, which can be represented as 
\begin{equation}
K_e = K_s^{(1-\phi)}K_i^{S\phi}K_w^{(1-S)\phi},\label{eqn:effk}
\end{equation} 
for the thermal conductivities of the fringe constituents, sediment $K_s$, ice $K_i$, and water $K_w$ \citep{Cla1995}. Using the divergence theorem, we can write equation \eqref{eqn:enthalpy1} as
\begin{eqnarray}
\pd{}{t}\left[ (1-\phi)\rho_s h_s + \phi S\rho_{i}h_{i} + \phi (1-S) \rho_{w}h_{w}\right] = \hspace{7cm} \nonumber \\ -\boldsymbol{\nabla}\cdot \left\{\phi S \rho_i h_i \boldsymbol{V} + \phi \left( 1- S\right) \rho_{w}h_{w}\boldsymbol{U} + \left( 1-  \phi \right)\rho_{s}h_{s}\boldsymbol{V}_s  \right\} + \boldsymbol{\nabla}\cdot\left( K_e \boldsymbol{\nabla}T\right),\label{eqn:consenergy}
\end{eqnarray}
We define the difference between the water flux $\boldsymbol{U}$ and the heave rate $\boldsymbol{V}$ as $\boldsymbol{u}$, i.e.  
\begin{equation}
\boldsymbol{u} = \boldsymbol{U}-\boldsymbol{V},
\end{equation}
which will typically be small as it is the flow of water that allows for heave. Water flow is given by Darcy's law \eqref{eqn:darcyfirst} as 
\begin{equation}
\phi (1-S) \boldsymbol{u} = -\phi (1-S) \boldsymbol{V} - \frac{k(S)}{\mu}\left(\pd{p_w}{z}+\rho_w g\right).
\end{equation}
Thus, we combine the mass conservation equations \eqref{eqn:icecons} and \eqref{eqn:watercons} as
\begin{eqnarray}
\pd{}{t}\left[\phi S\rho_i + \phi (1-S)\rho_{w}\right] + \boldsymbol{\nabla}\cdot \left\{\left[\phi S\rho_i + \phi (1-S)\rho_{w}\right]\boldsymbol{V}\right\} \hspace{2cm} \nonumber \\
+ \boldsymbol{\nabla}\cdot\left[\rho_w \phi \left(1-S\right) \boldsymbol{u}\right] = 0,\label{eqn:massconscombine}
\end{eqnarray}
which allows us to define the total ice and water $W$ \citep[e.g.][]{Mey2017c} that is given by
\begin{eqnarray}
W = \phi S\rho_i + \phi (1-S) \rho_{w},
\end{eqnarray}
so that in the rigid-ice limit with $\boldsymbol{\nabla} \cdot \boldsymbol{V}=0$ where the heave rate is spatially independent, mass conservation \eqref{eqn:massconscombine} can be written succinctly as 
\begin{eqnarray}
\pd{W}{t}+ \boldsymbol{V}\cdot\boldsymbol{\nabla}W + \boldsymbol{\nabla}\cdot \left[ \rho_w \phi \left(1-S\right) \boldsymbol{u}\right] = 0.\label{eqn:massconscombineW}
\end{eqnarray}

\begin{figure}[!ht]
    \includegraphics[width=0.5\linewidth]{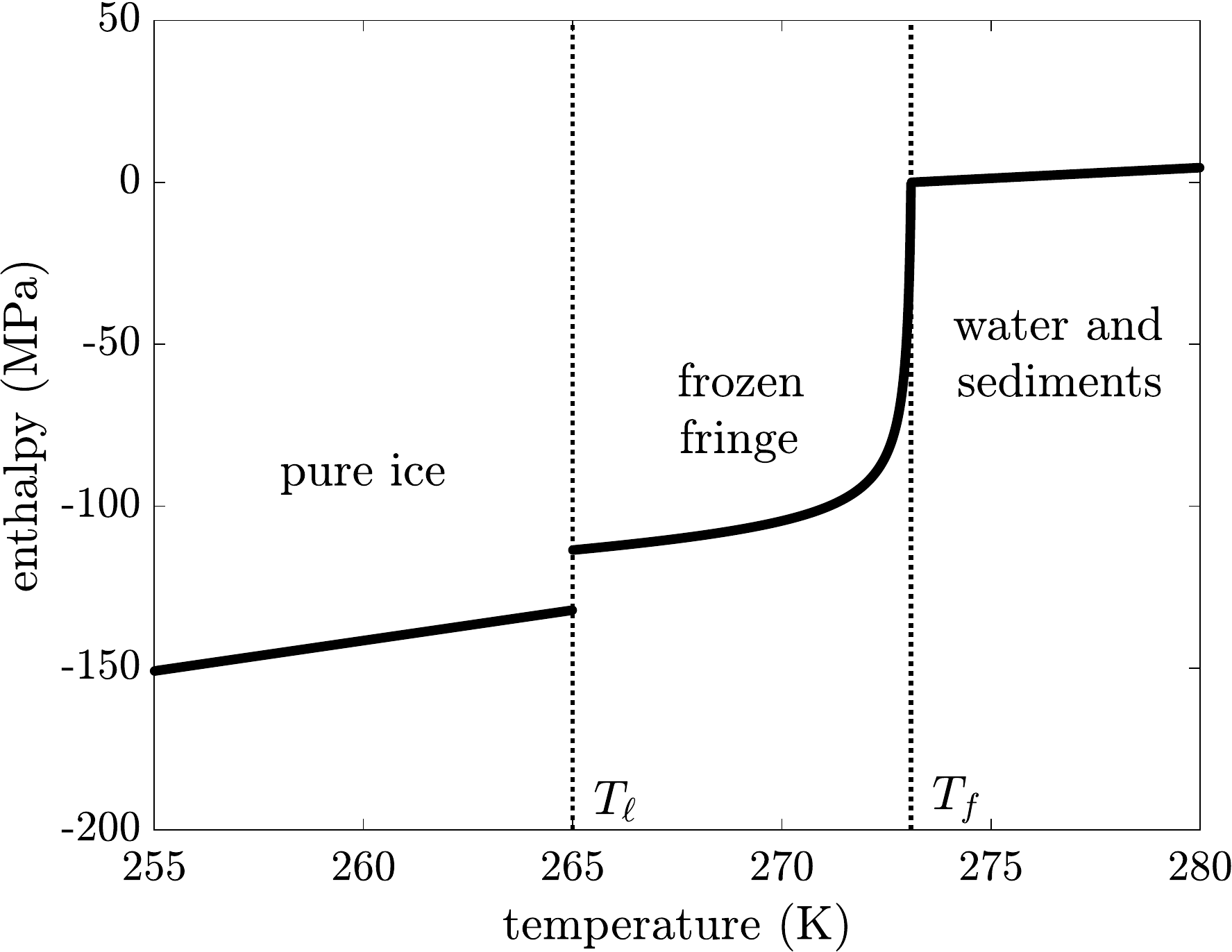}
     \includegraphics[width=0.5\linewidth]{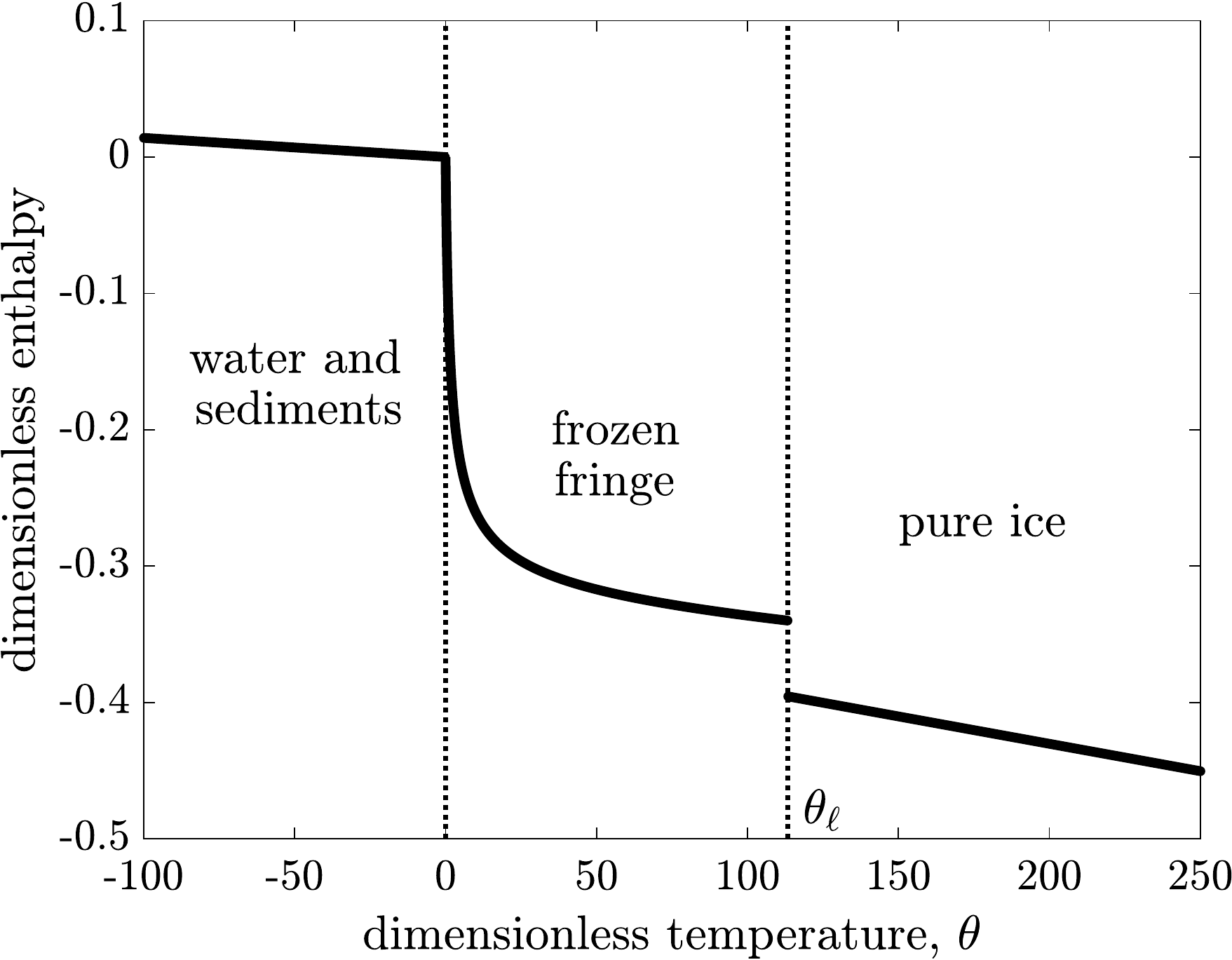}
    \caption{Schematic for the enthalpy below an ice lens: (a) Temperatures below $T_\ell$ occur above the lowest ice lens; within the fringe, the temperature is tied to the ice saturation curve; and below the fringe, the water and sediment temperatures rise above $T_f$. (b) Nondimensional version of (a) showing that the enthalpy is zero at $\theta=0$ and negative nondimensional temperatures correspond to unfrozen water with positive enthalpy.}
    \label{fig:enthalpyphasediagram}
\end{figure}

Similarly, for conservation of energy we can define the total enthalpy $H$ as
\begin{eqnarray}
H = H_0+\phi S\rho_{i}h_{i} + \phi (1-S) \rho_{w}h_{w},\label{eqn:bigHdef}
\end{eqnarray}
which is specified up to a constant, reference enthalpy $H_0$, which we choose to make $H=0$ at $T=T_f$ Thus, equation \eqref{eqn:consenergy} reduces to
\begin{eqnarray}
\pd{H}{t}+\bs{V}\cdot\bs{\nabla}H +\pd{}{t}\left[ (1-\phi)\rho_s h_s\right] + \bs{\nabla}\cdot\left[\phi \left( 1- S\right) \rho_{w}h_{w}\bs{u} + \left( 1- \phi\right) \rho_{s}h_{s}\bs{V}_s \right] = \bs{\nabla}\cdot\left( K_e \bs{\nabla}T\right).\label{eqn:consenergyH}
\end{eqnarray}
Expanding out the total enthalpy \eqref{eqn:bigHdef} and using the fact that the specific latent heat $\mathscr{L}$ is equal to the difference between the liquid and ice specific enthalpies, i.e. $\mathscr{L} = h_w - h_i$, we find that 
\begin{eqnarray}
H = H_0+Wh_{i} +\rho_w \mathscr{L} \phi (1-S),
\end{eqnarray}
which is a sum of sensible and latent heat contributions to energy. We leave equation \eqref{eqn:consenergyH} in enthalpy form to facilitate the description of the numerical method. For an incompressible medium, the specific enthalpy $h_\alpha$ is equivalent to a change in temperature, i.e. $dh_\alpha = c_\alpha dT$, where $c_\alpha$ is the specific heat capacity with $\alpha$ representing ice, liquid, or sediments. Thus, the enthalpy can be related to temperature as
\begin{equation}
H = \left \{ 
\begin{array}{ccc}
\rho_i c_i \left(T - T_f\right)-\rho_w \mathscr{L} & T <T_\ell & \mbox{(pure ice)}\\
W c_i \left(T - T_f\right)-\rho_w \mathscr{L} \phi S & T_\ell \leq T < T_f & \mbox{(frozen fringe)}\\ 
\rho_w c_i \phi \left(T - T_f\right)& T_f  \leq T & \mbox{(water and sediments)}
\end{array}\right. ,
\end{equation}
which we show schematically in figure \ref{fig:enthalpyphasediagram}. As we describe in the next section, the ice saturation in the fringe is a function of the temperature.


\subsection{Constitutive relations for saturation and permeability}
The ice saturation $S$ and, therefore, the permeability $k$ of the frozen fringe depend on the local thermodynamics with the pressure difference between the phases controlled by the Gibbs-Thomson effect and interfacial premelting \citep{And2004,Han2010,Rem2012}. Here we use the generalized Clausius-Claperyon relation to specify the ice saturation $S_i$ and permeability $k$ as functions of the local difference between ice pressure and water pressure. That is, we generalize the relationships used by \citet{Rem2007,Rem2008} and define the function $\Xi$ as the ratio of the critical effective stress to the local pressure difference $p_i-p_w$, i.e. 
\begin{equation}
\Xi = \frac{N_c}{p_i - p_w} = \frac{T_m - T_f}{T_m - T(z) + \frac{\left(\rho_w - \rho_i\right)T_m}{\rho_i \rho_w \mathscr{L}}\left(N_{\textrm{loc}} + p_w(z_f) - p_w(z) \right)}.
\label{xifun}
\end{equation}
This reduces to the expression used by \citet{Rem2007,Rem2008} if we ignore the density difference between ice and water. For now, we proceed with the definition in equation \eqref{xifun} and write the ice saturation and permeability as
\begin{eqnarray}
S &=& 1 - \Xi^{\beta},\\ 
k &=& k_0 \left(1 - S\right)^{\alpha/\beta}, 
\end{eqnarray}
where the empirical exponents are $\beta>0$ and $\alpha>\beta$ \citep[typically $\alpha>1$;][]{Rem2008}.

\subsection{Boundary conditions}
Now that we have specified the governing equations for enthalpy $H$ and total water $W$, we describe the boundary conditions. At the top of the lowest fringe, i.e. $z=z_{\ell}$, a finite jump in saturation can occur as the ice lens is fully occupied by ice ($\phi=1$ and $S=1$) whereas ice only partially saturates the interstices in the underlying fringe ($\phi<1$ and $S<1$). Integrating mass and energy conservation across the jump at the lowest ice lens boundary gives the following conditions
\begin{eqnarray}
\left[W \bs{V} + \rho_w\phi \left(1-S\right) \bs{u}\right]_{-}^{+}\cdot \bs{n}  &=& 0~~~\mbox{across}~~~z=z_{\ell},\\
\left[-K_e \bs{\nabla}T + H\bs{V} + \phi \left( 1- S\right) \rho_{w}h_{w}\bs{u} + \left( 1- \phi\right) \rho_{s}h_{s}\bs{V}_s\right]_{-}^{+}\cdot\bs{n}  &=& 0~~~\mbox{across}~~~z=z_{\ell}.
\end{eqnarray}
Taking the heat flux into the ice lens and material above as $q$, we can simplify these conditions to
\begin{eqnarray}
\left[W \bs{V} + \rho_w\phi \left(1-S\right) \bs{u}\right]_{-}\cdot\bs{n} &=& \rho_i \bs{V}\cdot\bs{n}~~~\mbox{at}~~~z=z_{\ell}\\
\left[-K_e \bs{\nabla}T + H\bs{V} + \phi \left( 1- S\right) \rho_{w}h_{w}\bs{u} + \left( 1- \phi\right) \rho_{s}h_{s}\bs{V}_s\right]_{-}\cdot\bs{n} &=& \bs{q}\cdot\bs{n}~~~\mbox{at}~~~z=z_{\ell},
\end{eqnarray}
where mass conservation implies that at the top of the fringe, all of the liquid water must freeze onto the lowest ice lens, and conservation of energy implies that all the heat that enters the fringe at the bottom must leave through the top. At the bottom of the fringe (or at any point below the fringe), we impose a conductive heat flux $q$ which includes contributions from geothermal heat and friction from sliding, i.e.
\begin{equation}
\left.-K_e \bs{\nabla}T\right |^{+}\cdot\bs{n} = \bs{q}\cdot\bs{n} ~~~\mbox{on}~~~z=z_{f},
\end{equation}
which is the full heat flux as $H=0$ at the base of the fringe (i.e. where we have chosen $H=H_0$ such that $H=0$ when $T=T_f$). The ice saturation transitions smoothly from $0<S<1$ within the fringe to $S=0$ below the fringe and, therefore, no jump condition is required at $z=z_f$. The water pressure $p_w$ at the bottom of the fringe is set by the effective pressure $N$ and is given by 
\begin{equation}
p_w = \sigma_n - N~~~\mbox{on}~~~z=z_{f}.
\end{equation}

\subsection{Nondimensionalisation}
We now scale our model to find the dominant physical balances. For example, we write $t = [t]t^{*}$ for the time $t$, where $[t]$ is the scale and $t^{*}$ is the nondimensional variable. Proceeding in this way, we write all variables as
\begin{eqnarray}
N = [N]N^{*},~~~p_w = \sigma_n - N + [N]p_w^{*},~~~T = T_f - [T]\theta,~~~K_e = K_i K^{*},~~~V = [V]V^{*} \nonumber\\
z = [z]z^{*},~~~t= [t]t^{*},~~~k = [k]k^{*},~~~H = \rho_w \mathscr{L} H^{*},~~~W = \rho_w W^{*},~~~h = c_i [T] h^{*},\nonumber
\end{eqnarray}
and choose the scales for the variables based on the expected physical balances. 

A scale for the effective pressure within the fringe is the threshold entry pressure, i.e. $[N] = N_c$, and a scale for the heat flux into the fringe is the geothermal heat flux, i.e. $[q]=\bs{q}\cdot\bs{n}$. We choose the temperature scale to be the temperature difference implied by premelting, i.e. 
\begin{equation}
[N] = \frac{\rho_i \mathscr{L} [T]}{T_m}\longrightarrow [T] = \frac{T_m [N]}{\rho_i \mathscr{L}}.
\end{equation}
A scale for the vertical distance $[z]$ comes from the heat flux scale, i.e.
\begin{equation}
[q] = K_i\frac{[T]}{[z]} \longrightarrow [z] = K_i\frac{[T]}{[q]}.
\end{equation}
The rate of heaving is determined by water percolation, so we choose
\begin{equation}
[V] = \frac{[k][N]}{\mu[z]},
\end{equation} 
and time can be scaled for the solidification as
\begin{equation}
\frac{\rho_w \mathscr{L}}{[t]}=K_i \frac{[T]}{[z]^2} \longrightarrow [t]=\frac{\rho_w \mathscr{L}[z]^2}{K_i [T]}.
\end{equation}
For the permeability, we choose the scale to be the prefactor as 
\begin{equation}
[k] = k_0.
\end{equation}

We also define the dimensionless variables
\begin{eqnarray}\delta = 1 - \frac{\rho_i}{\rho_w},~~~\nu = \frac{\rho_s}{\rho_w},~~~\mathtt{Pe} = \frac{[V][t]}{[z]},~~~\mathtt{Gr} = \frac{\rho_w g [z]}{[N]},~~~\mathtt{St} = \frac{\mathscr{L}}{c_i [T]},\end{eqnarray}
where $\delta$ is the scaled density difference, $\nu$ is the ratio of the sediment density to water density, $\mathtt{Pe}$ is the P\'{e}clet number, $\mathtt{Gr}$ is the ratio of gravitational hydrostatic pressure to infiltration pressure, and $\mathtt{St}$ is the Stefan number. 

\subsubsection{Full model}
We now write the model in nondimensional variables and for concision, we drop the asterisks. Rewriting equation \eqref{esfullforce_T}, the force balance across the fringe is
\begin{eqnarray}
N =1 + \mathtt{Gr}\int_{z_f}^{z_{\ell}}{\left[\nu (1-\phi) + \phi\left(1-S\right)\right]dz'}+ \int_{z_f}^{z_{\ell}}{\left( 1 - \phi S\right) \left[\pd{\theta}{z'}+ \left( 1-\delta\right) \pd{p_w}{z'}\right]dz'},
\end{eqnarray}
if a fringe exists, or the effective pressure is constrained by $N<1$ if there is not a fringe. The local effective pressure in the fringe is 
\begin{eqnarray}
N_{\textrm{loc}}(z) = N - \left\{\mathtt{Gr}\int_{z_f}^{z}{\left[\nu (1-\phi) + \phi\left(1-S\right)\right]dz'} - \int_{z_f}^{z}{\phi S \pd{\theta}{z'} dz'} +  \right. \hspace{3.75cm} \nonumber  \\ \left. \phi S\left[1 + \theta+ \delta\left(N - p_w(z)\right)\right] + \int_{z_f}^{z}{\left[1-\left( 1-\delta \right) \phi S \right]\pd{p_w}{z'}dz'} \right\}.\hspace{0cm}\label{localN_T_nd}
\end{eqnarray}

The total water in the fringe is
\begin{eqnarray}
W &=& \phi\left(1-\delta S\right),
\end{eqnarray}
and the enthalpy is
\begin{eqnarray}
H &=& \frac{1}{\mathtt{St}}Wh_{i} - \phi S = \left \{ 
\begin{array}{ccc}
-(1-\delta)(\theta/\mathtt{St})-1 & \theta>0~\&~\phi=1 & \mbox{(pure ice)}\\
-\phi (1-\delta S)(\theta/\mathtt{St})-\phi S & \theta>0& \mbox{(frozen fringe)}\\ 
-\phi( \theta/\mathtt{St})& \theta \leq 0 & \mbox{(water and sediments)}
\end{array}\right.
\end{eqnarray}
Nondimensionalising equation \eqref{eqn:consenergyH} results in
\begin{eqnarray}
\pd{H}{t}+\mathtt{Pe}\bs{V}\cdot\bs{\nabla}H +\frac{\nu}{\mathtt{St}} \pd{}{t}\left[ (1-\phi) h_s\right] + \frac{\mathtt{Pe}}{\mathtt{St}} \bs{\nabla}\cdot\left[\phi \left( 1- S\right) h_{w}\bs{u} + \nu (1-\phi) h_s \bs{V}_s\right] = -\bs{\nabla}\cdot\left( K \bs{\nabla}\theta\right),\label{eqn:consenergyHnd}
\end{eqnarray}
The scaled total water equation is given by 
\begin{eqnarray}
\pd{W}{t}+ \mathtt{Pe}\bs{V}\cdot\bs{\nabla}W + \mathtt{Pe}\bs{\nabla}\cdot \left[ \phi \left(1-S\right) \bs{u}\right] = 0,\label{eqn:massconscombineWfull}
\end{eqnarray}
which depends on both the flow of water $\bs{u}$ and the rate of heave $\bs{V}$.


The constitutive laws for permeability and saturation are written nondimensionally as 
\begin{eqnarray}
k &=& \Xi^{\alpha},\\
S &=& 1 - \Xi^{\beta},
\end{eqnarray}
where
\begin{equation}
\Xi = \frac{1}{1+ \theta + \delta\left[N_{\textrm{loc}} - p_w(z) \right]}.
\label{xifun_nd}
\end{equation}

Finally, the nondimensional boundary conditions are given as 
\begin{eqnarray}
\left[W\bs{V} +\phi \left(1-S\right) \bs{u}\right]_{-}\cdot\bs{n} &=& \left(1-\delta\right) \bs{V}\cdot\bs{n}~~~\mbox{at}~~~z=z_{\ell},\label{bcatl1}\\
\left[K\bs{\nabla}\theta + \mathtt{Pe}H\bs{V} + \frac{\mathtt{Pe}}{\mathtt{St}}\phi \left( 1- S\right) h_{w}\bs{u} + \frac{\nu\mathtt{Pe}}{\mathtt{St}}\left( 1- \phi\right) h_{s}\bs{V}_s\right]_{-}\cdot \bs{n} &=& 1~~~\mbox{at}~~~z=z_{\ell},\label{bcatl2}\\
K\left.\bs{\nabla}\theta\right|^{+}\cdot\bs{n} &=& 1 ~~~\mbox{below}~~~z=z_{f},\\
p_w &=& 0~~~\mbox{on}~~~z=z_{f}.
\end{eqnarray}
%

\begin{table}
\centering
\begin{tabular}{llllllll}
\hline
\multicolumn{4}{c}{fringe parameters} & \multicolumn{2}{l}{variable scales} & \multicolumn{2}{l}{nondimensional values}\\ 
\hline
$\rho_i$ & 917 kg m$^{-3}$ & $\gamma$ & 0.034 J m$^{-2}$ & $[T]$ & 0.061 K & $\delta$ & 0.083\\
$\rho_w$ & 1000 kg m$^{-3}$ & $\mu$ & $1.8\times10^{-3}$ Pa s & $[q]$ & $0.070$ W m$^{-2}$ & $\nu$ & 2.5\\
$\rho_s$ & 2500 kg m$^{-3}$ &  $r_p$ & 10$^{-6}$ m & $[z]$ & 1.8 m & $\mathtt{Pe}$ & 0.91\\
$c_i$ & 2050 m$^{2}$ s$^{-2}$ K$^{-1}$ & $\alpha$ & 3.1 & $[V]$ & 6.5 mm yr$^{-1}$ &$\mathtt{Gr}$ & 0.26 \\ 
$c_w$ & 4200 m$^{2}$ s$^{-2}$ K$^{-1}$ & $\beta$ & 0.53 & $[t]$ &  250 yr & $\mathtt{St}$ & 2700\\
$c_s$ & 800 m$^{2}$ s$^{-2}$ K$^{-1}$ &  $N$ & 100 kPa & $[k]$ & $10^{-17}$ m$^{2}$ & \\ 
$K_i$ & 2.1 kg m s$^{-3}$ K$^{-1}$  &  $\sigma_n$ & 1000 kPa& $[N]$ & 68 kPa & \\
$K_w$ & 0.56 kg m s$^{-3}$ K$^{-1}$  & $\phi$ & 0.35 \\
$K_s$ & 4.0 kg m s$^{-3}$ K$^{-1}$  & $k_0$ & $10^{-17}$ m$^{2}$ \\
$\mathscr{L}$ & $3.34\times10^5$ m$^{2}$ s$^{-2}$ &  $T_m$ & 273.15 K \\
$g$ & 9.80 m s$^{-2}$ &  $q$ & $0.070$ W m$^{-2}$ \\

\hline
\end{tabular}
\caption{Table of parameters for frozen fringe.}
\label{prms}
\end{table}

\subsubsection{Model reduction\label{reduction}}
Typical values for the nondimensional variables based on the parameters are given in table \ref{prms}. The scaled density difference $\delta$ is a small value and therefore it is reasonable to neglect terms that are multiplied by $\delta$ \citep{Rem2008}. Taking this limit, we find that the vertical force balance reduces to 
\begin{equation}
N = 1 + \mathtt{Gr}\int_{z_f}^{z_{\ell}}{\left[\nu (1-\phi) + \phi (1-S)\right]dz'} 
+ \int_{z_f}^{z_{\ell}}{\left( 1 - \phi S\right) \left[\pd{\theta}{z'}+  \pd{p_w}{z'}\right]dz'},\label{eqn:vfbr}
\end{equation}
unless $N<1$, in which case there is not a fringe. In the same way, the local effective pressure $N_{\textrm{loc}}(z)$ is
\begin{eqnarray}
N_{\textrm{loc}}(z) = N - \left\{\mathtt{Gr}\int_{z_f}^{z}{\left[\nu (1-\phi) + \phi (1-S)\right]dz'} - \int_{z_f}^{z}{\phi S \pd{\theta}{z'} dz'} +  \right. \hspace{3.5cm} \nonumber  \\ \left. \phi S\left[1 + \theta\right] + \int_{z_f}^{z}{\left(1-\phi S\right) \pd{p}{z'}dz'}  \right\}.\label{localN_T_nd2}
\end{eqnarray}

Now since the Stefan number $\mathtt{St}$ is large, the sensible heat contributions to the enthalpy $H$ within the fringe can be ignored. Thus, we have that
\begin{eqnarray}
H &=& \left \{ 
\begin{array}{ccc}
-(\theta/\mathtt{St})-1 & \theta>0~\&~\phi=1 & \mbox{(pure ice)}\\
-\phi S & \theta>0& \mbox{(frozen fringe)}\\ 
-\phi( \theta/\mathtt{St})& \theta \leq 0 & \mbox{(water and sediments)}
\end{array}\right.
\end{eqnarray}
to leading order in $1/\mathtt{St}$ in the fringe. Only sensible heat terms persist in the pure ice and water and sediments and we retain the $1/\mathtt{St}$ dependence to meet flux boundary conditions. The enthalpy variation in the frozen fringe is tied to the temperature $\theta$ through the ice saturation $S$ as 
\begin{equation}
H = -\phi S = -\phi\left[1 - \left( 1+\theta\right)^{-\beta}\right],
\label{eqn:Hfringe}
\end{equation}
analogous to the liquidus condition in a mushy zone \citep{Wor2000}.

Given that frozen fringes are often much wider than thick, we now restrict our attention to one vertical dimension for conservation of mass and energy. Thus, in the same large Stefan number limit, the evolution equation for enthalpy is
\begin{eqnarray}
\pd{H}{t}+\mathtt{Pe}V\pd{H}{z} = -\pd{}{z}\left( K \pd{\theta}{z}\right),\label{eqn:consenergyHndreduced}
\end{eqnarray}
where we have neglected terms proportional to $\mathtt{Pe}/\mathtt{St}$ as well. In the limit $\delta=0$, the total water $W$ reduces to 
\begin{equation}
W = \phi,
\end{equation}
which is a constant, meaning that the pore space is entirely occupied by ice and water, yet there is no distinction in this limit due to the small density difference. Therefore, mass conservation implies
\begin{eqnarray}
V\pd{}{z}\left[ \phi \left(1-S\right)\right] = \pd{}{z}\left[k(S)\left(\pd{p_w}{z}+\mathtt{Gr}\right)\right].\label{eqn:Wphi}
\end{eqnarray}

The boundary conditions for mass, momentum, and energy conservation reduce to
\begin{eqnarray}
\left[\phi S V +k(S)\left(\pd{p_w}{z}+\mathtt{Gr}\right)\right]_{-} &=& V~~~\mbox{at}~~~z=z_{\ell},\label{eqn:bcatl1r}\\
\left[K\pd{\theta}{z} + \mathtt{Pe}VH \right]_{-} &=& 1~~~\mbox{at}~~~z=z_{\ell},\label{eqn:bcatl2r}\\
K\left.\pd{\theta}{z}\right|^{+} &=& 1 ~~~\mbox{below}~~~z=z_{f},\label{eqn:bcatl3r}\\
p_w &=& 0~~~\mbox{on}~~~z=z_{f}.
\end{eqnarray} 

We now integrate equation \eqref{eqn:Wphi} and impose the boundary condition \eqref{eqn:bcatl1r}, which implies that the water pressure gradient is
\begin{equation}
\pd{p_w}{z}= -\mathtt{Gr}  -\frac{V}{k}\left(1-\phi S\right).
\end{equation}
We can now insert this water pressure gradient into the vertical force balance \eqref{eqn:vfbr} to find that the heave rate $V$ is given by
\begin{equation}
V = \frac{1 -N+ \mathtt{Gr}\left(\nu-1\right) (1-\phi) \left(z_{\ell}- z_f\right) 
+ \int_{z_f}^{z_{\ell}}{\left( 1 - \phi S\right)\pd{\theta}{z'}dz'}}{\int_{z_f}^{z_{\ell}}{\frac{(1-\phi S)^2}{k}dz'}}, \label{eqn:hr_nd}
\end{equation}
as shown previously by \citet{Rem2008}. This prescription of the heave rate is determined by force balance as well as conservation of mass and requires integrating the temperature field $\theta$ and the attendant saturation $S$. Thus, we can summarise our full model for the transient evolution of a frozen fringe as: enthalpy evolution \eqref{eqn:consenergyHndreduced} with heave rate \eqref{eqn:hr_nd} subject to boundary conditions \eqref{eqn:bcatl2r} and \eqref{eqn:bcatl3r}.

\subsection{Enthalpy numerical method}
We write equation \eqref{eqn:consenergyHndreduced} in conservative form, defining the flux as the sum of the advective and diffusive components. We then discretise the conserved fluxes in space using a finite volume method implemented in python. In this numerical method, we divide the domain into cells and each variable is constant within a cell whereas velocities and fluxes are evaluated at cell edges. For advection, we use an upwinding scheme where the advective fluxes on cell edges are given by the cell values `upwind', which is determined by the sign of the heave rate. We evolve explicitly equation \eqref{eqn:consenergyHndreduced} in time using \emph{solve\_ivp} and the method of lines in python. With these choices, the finite volume implementation is conservative, meaning that the flux transferred between cells respects conservation of energy and phase change. The code is included in the supplemental information as well as in a github repository (link/doi to be added in proofs). 

The two input parameters for the model are the effective pressure $N$ and the heave rate $V$. Thus, coupling the governing equation \eqref{eqn:consenergyHndreduced} with the heave rate equation \eqref{eqn:hr_nd}, this problem takes an integro-differential equation form, where at each timestep we integrate equation \eqref{eqn:hr_nd}. Rather than implementing the top boundary condition \eqref{eqn:bcatl2r} as a total flux, we apply a boundary condition to the diffusive part as 
\begin{equation}
K\pd{\theta}{z}  = 1 - \mathtt{Pe}V_{\textrm{input}}H ~~~\mbox{at}~~~z=z_{\ell},
\end{equation}
where $V_{\textrm{input}}$ is the input value. The steady state is diagnosed when the value of $V$ computed through equation \eqref{eqn:hr_nd} is equal to $V_{\textrm{input}}$. We study the steady state problem in more detail in the next section.  The outputs for the model are the frozen fringe thickness $h = z_{\ell}-z_f$, the temperature profile through the fringe and the ice saturation profile in the fringe, as well as the initiation, timing, and spacing of ice lenses.

\section{Results\label{sec:results}}

\subsection{Steady state frozen fringe}
To understand the development of a frozen fringe as well as the relationship between a free boundary representation and the enthalpy method, we start by considering a steady state frozen fringe with constant thermal conductivity. The problem is then: for a fixed location of the lowest ice lens $z_\ell$ (e.g. the glacier-sediment interface), a constant heave rate $V$, and a known effective pressure $N$, what is the steady state temperature profile $\theta(z)$ and fringe-front location $z_f$?

Conservation of energy in the fringe and the water-saturated region in front of it is given by
\begin{eqnarray}
     \mathtt{Pe} V\td{H}{z} &=& -\td{^2\theta}{z^2}~~~(z_{f}<z<z_{\ell}),\label{eqn:uppernethalpy}\\
     \td{^2\theta}{z^2}&=&0~~~~~~~(0<z<z_f),\label{eqn:lowertemp}
\end{eqnarray}
where the enthalpy $H$ in the fringe is given by \eqref{eqn:Hfringe} as $H=-\phi S$.

The boundary conditions are
\begin{eqnarray}
      \td{\theta}{z} + \mathtt{Pe} V H &=& 1~~~\mbox{on}~~~z=z_{\ell},\label{eqn:ulhb}\\
     \td{\theta}{z} &=& 1~~~\mbox{on}~~~z=0,\label{eqn:as_ghfbc}
\end{eqnarray}
with the internal conditions
\begin{eqnarray}
     \theta = 0~~~\mbox{on}~~~z=z_f,\label{eqn:as_ic1}\\
     \left[\td{\theta}{z}\right]^{+}_{-} = 0~~~\mbox{at}~~~z=z_f.\label{eqn:as_ic2}
\end{eqnarray}

We start by deriving the temperature profile for the region below the fringe. By integrating equation \eqref{eqn:lowertemp}, we have
\begin{equation}
    \theta = z - z_f~~~(0\leq z \leq z_f),
\end{equation}
which satisfies the geothermal heat flux boundary condition \eqref{eqn:as_ghfbc} and the scaled temperature goes to zero at the bottom of the fringe to satisfy the internal condition \eqref{eqn:as_ic1}.

Now in the fringe, we integrate equation \eqref{eqn:uppernethalpy} once and apply the boundary condition \eqref{eqn:ulhb}, which results in
\begin{equation}
\td{\theta}{z} = 1+\mathtt{Pe}V\phi \left[1 - \left( 1+\theta\right)^{-\beta}\right]\label{eqn:simplethetaode},
\end{equation}
which is the governing ordinary differential equation for the temperature profile in the fringe, subject to the boundary condition $\theta=0$ at $z=z_f$. Thus, for a given heave rate $V$ and effective pressure $N$, the temperature profile is specified by \eqref{eqn:simplethetaode}. The only piece of information that is missing, is the location of the bottom of the fringe $z_f$, which we determine through force balance \eqref{eqn:hr_nd}. To solve for $\theta$ and $z_f$, we numerically integrate \eqref{eqn:simplethetaode} for the temperature, insert the solution into the force balance \eqref{eqn:hr_nd}, and find the fringe front $z_f$ using a root-finding algorithm (i.e. similar to a shooting method). 

\begin{figure}
    \includegraphics[width=0.5\linewidth]{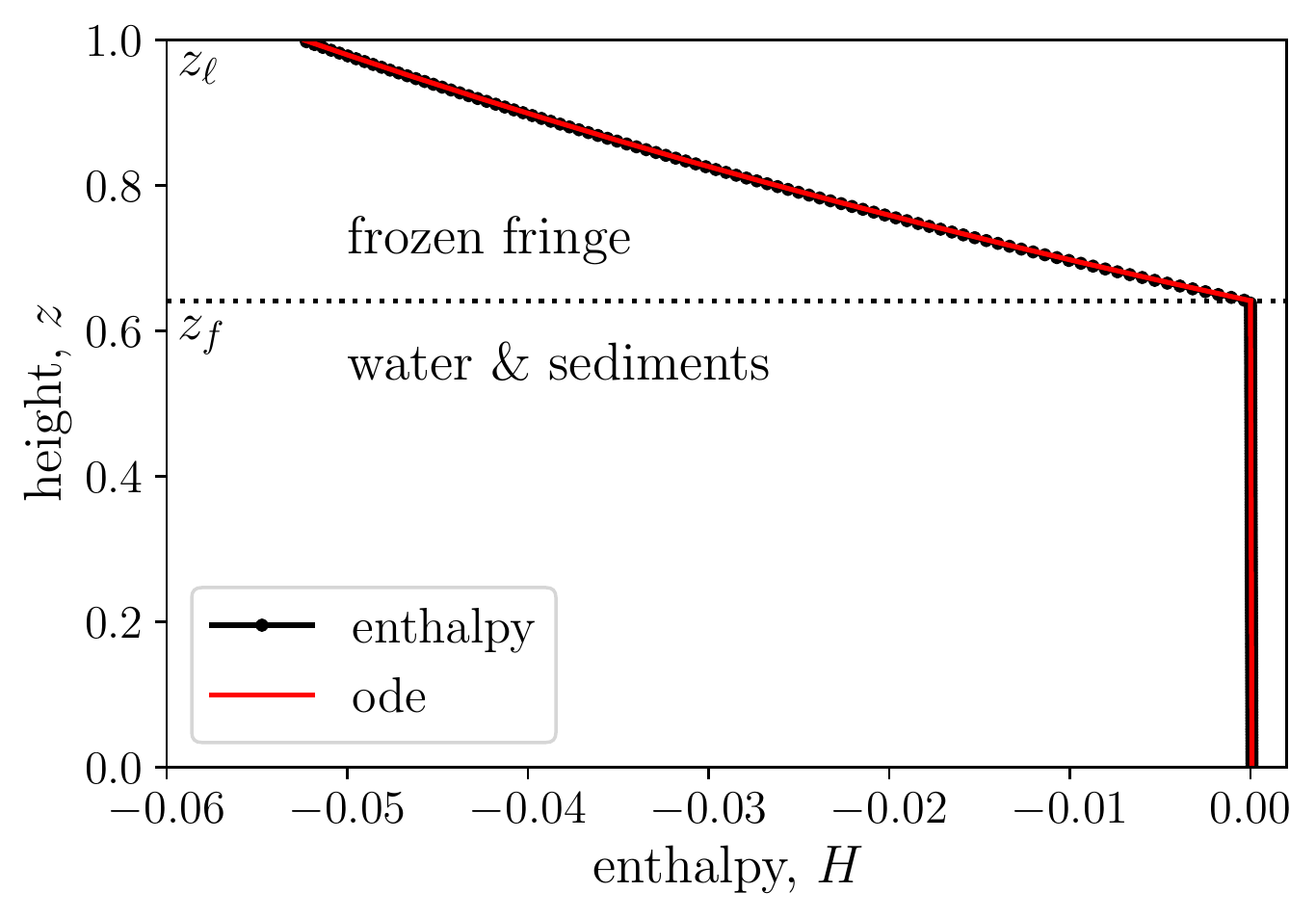}
    \includegraphics[width=0.51\linewidth]{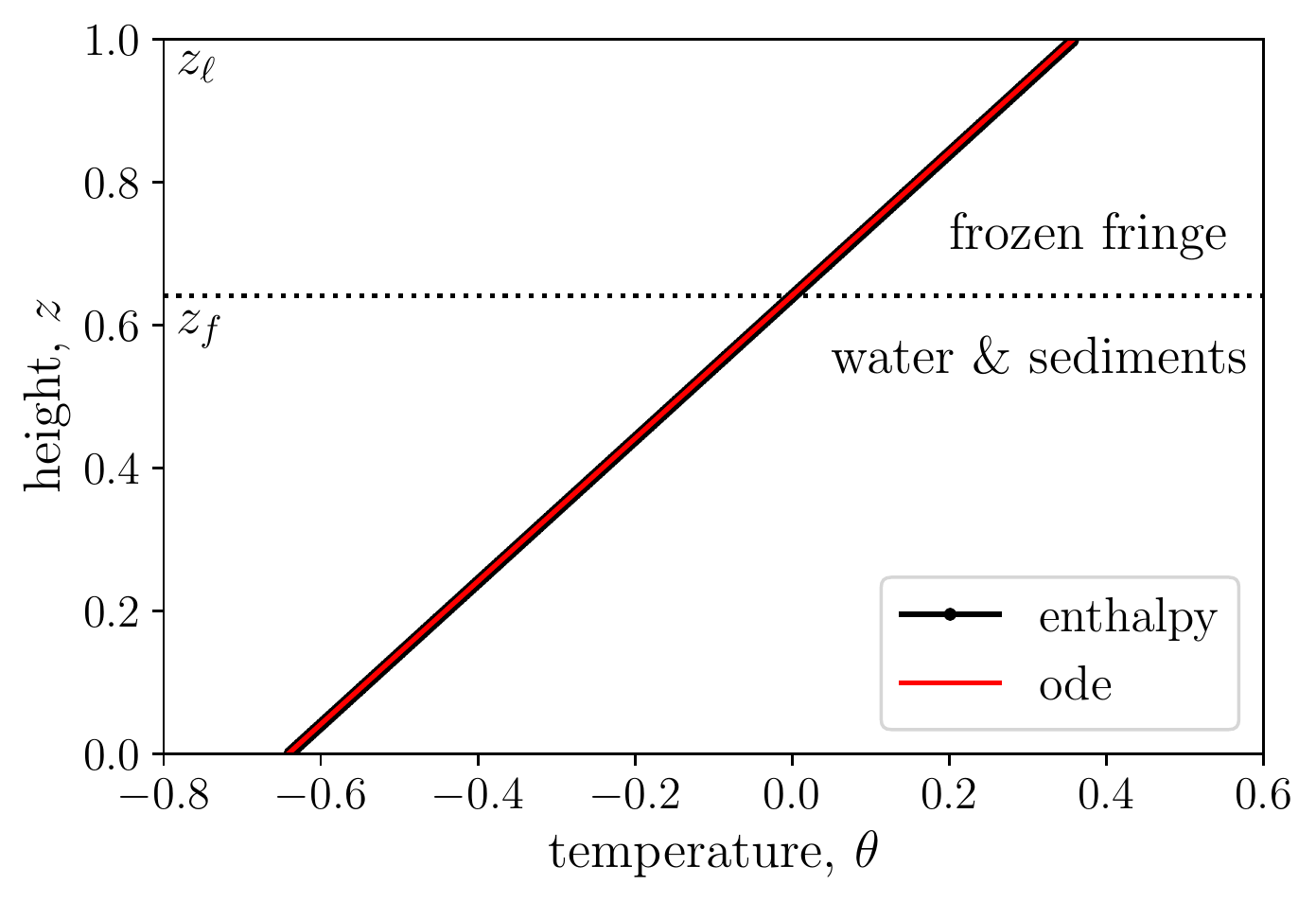}
    \caption{Comparison between the finite volume enthalpy method and the steady state ode solution (nondimensional parameters $V=-0.055$ and $N=1.5$). (left) enthalpy $H$ with height $z$ below the lowest ice lens at $z_{\ell}=1$. The frozen fringe extends from $z_f = 0.64$ to $z_{\ell}=1$, with water and sediments below. The nondimensional fringe thickness is $h = z_{\ell}-z_f = 0.36$. The enthalpy in the water and sediments portion is close to zero throughout the depth. (right) temperature $\theta$ as a function of height $z$ as calculated from the enthalpy.}
    \label{fig:enthalpyodecomparison}
\end{figure}

The result of this numerical procedure with the parameters given in table \ref{prms} is shown as the red line in figure \ref{fig:enthalpyodecomparison}. The black line shows the solution to the same problem using the enthalpy method, where the domain runs from $z=0$ to $z=z_{\ell}$. Here we set $z_{\ell}=1$. We find the steady state through a relaxation method, i.e. we integrate equation \eqref{eqn:consenergyHndreduced} in time until the difference between the computed heave rate and the target value is less than $10^{-3}$. The enthalpy in the frozen fringe is negative, taking on its smallest value at the base of the lowest ice lens $z_\ell$ and rising monotonically up to zero at the base of the fringe $z_f$. The enthalpy is positive in the water--saturated sediments, yet is very small due to the large Stefan number $\mathtt{St}$. The nondimensional temperature $\theta$ is shown in the right panel of figure \ref{fig:enthalpyodecomparison}. In line with $T=T_f-[T]\theta$, we see that $\theta$ is positive in the fringe and negative below. The temperature profile is close to linear, which makes sense given that the heave rate $V$ is small, and is exactly linear in the thermodynamically balanced case where $V=0$ and when the thermal conductivity is constant. At the bottom of the fringe $\theta=0$ and the location $z_f$ is determined through force balance. The nondimensional fringe thickness is given as $h=z_\ell - z_f$ and we find $h=0.36$ for the parameters in figure \ref{fig:enthalpyodecomparison}.

\begin{figure}
\begin{overpic}[width=0.49\linewidth]{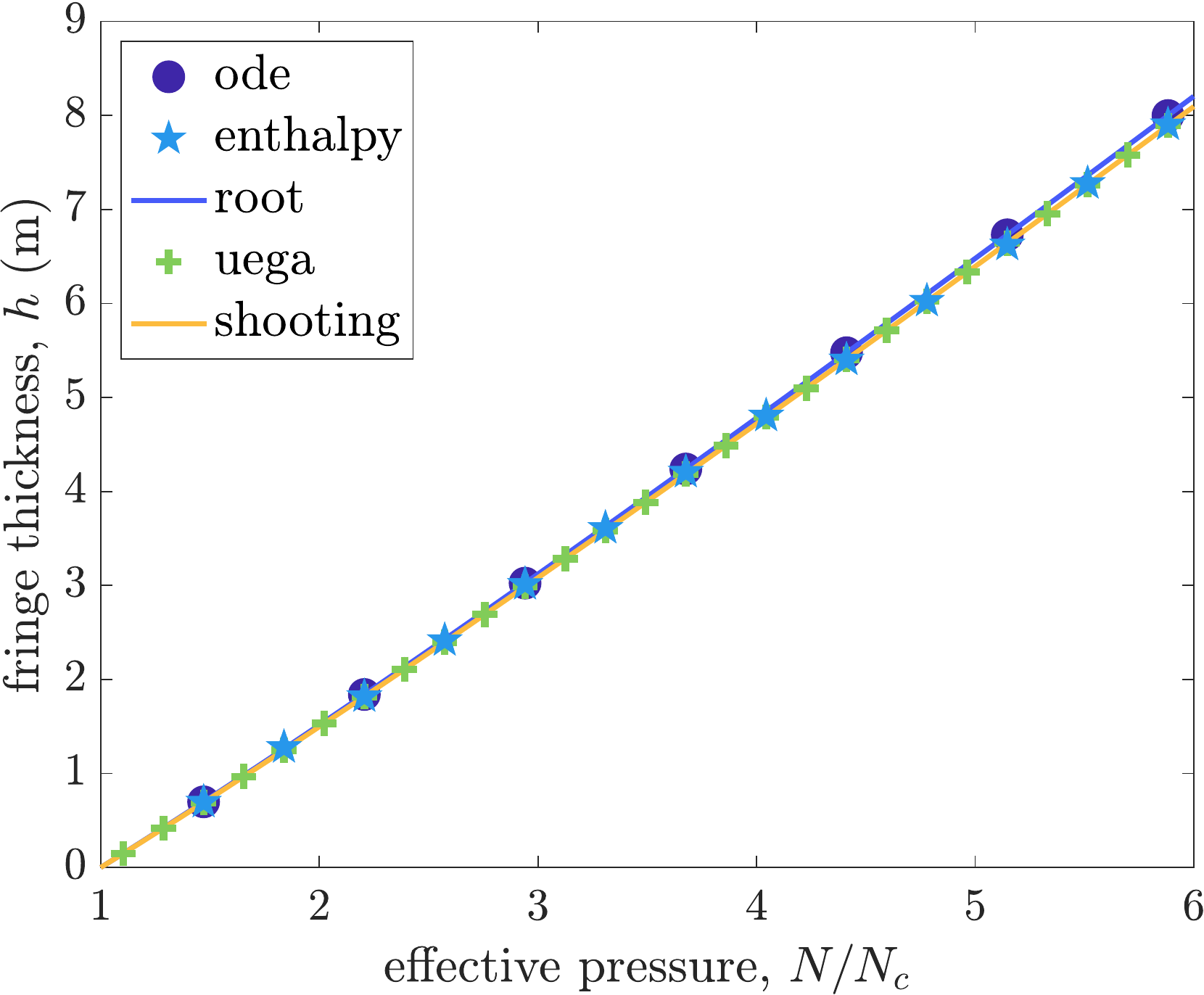}\put(40,74){balanced, $V=0$}
\end{overpic}
\begin{overpic}[width=0.49\linewidth]{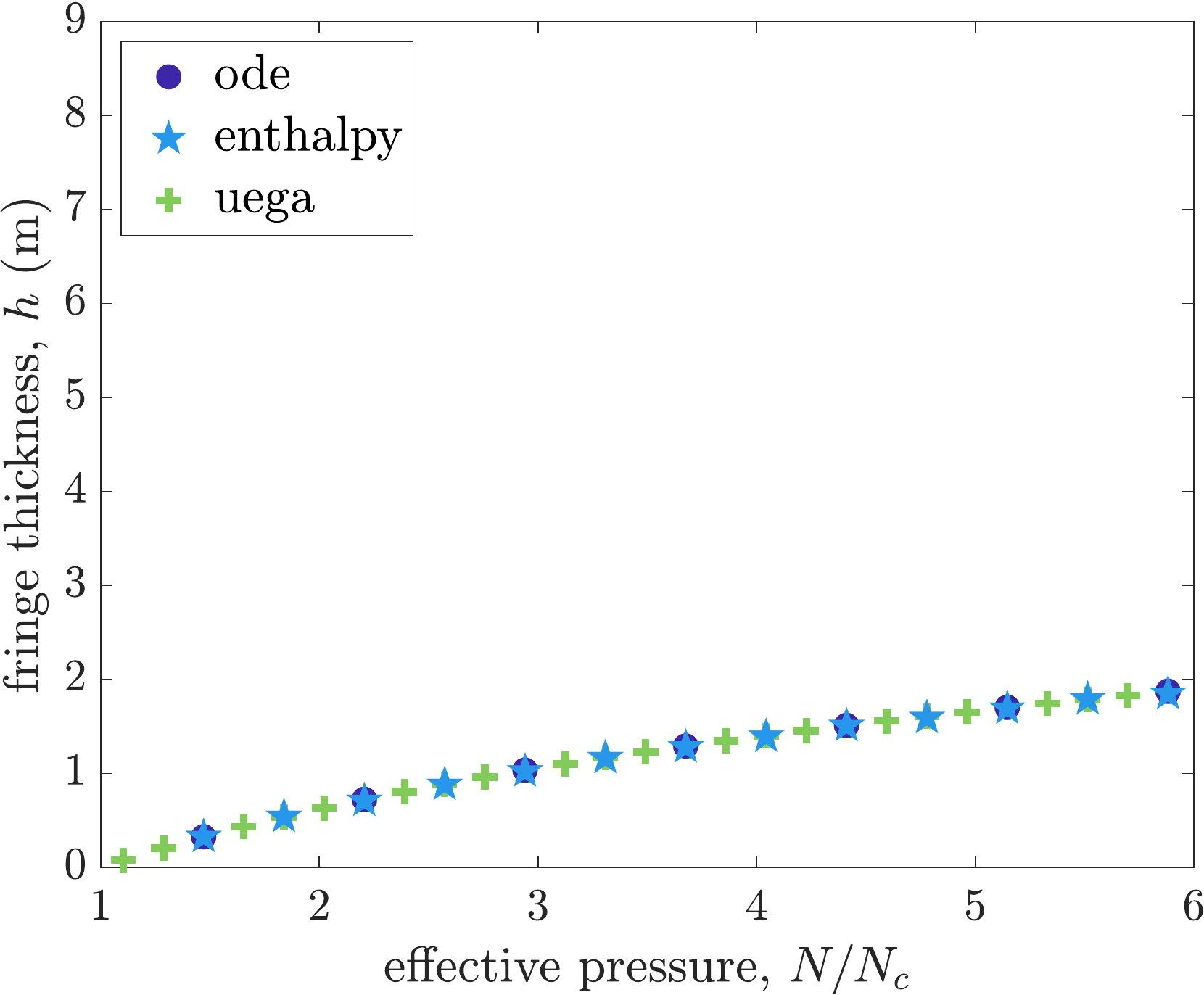}\put(40,74){melting, $V=-1.1$}
\end{overpic}
    \caption{Compilation of benchmark solutions for dimensional fringe thickness $h$ for two nondimensional heave rates (left) $V=0$ and (right) $V=-q/(\rho_i \mathscr{L} [V]) = -1.1$. The `uega' and `shooting' are solution methods from \citet{Rem2008}. The `enthalpy' solution is from the finite volume method, `ode' is the steady state ode solution, and `root' is the semi-analytical root-finding method. All methods give the same result.}
    \label{fig:hNcomparison}
\end{figure}

In figure \ref{fig:hNcomparison}, we show the dimensional fringe thickness $h = [h](z_\ell - z_f)$ in meters as a function of the nondimensional effective pressure $N/N_c$  for balanced ($V=0$, left) and temperate melting ($V = -q/(\rho_i \mathscr{L} [V])$, right) thermodynamics using 5 different solution methods. The first two techniques are what we just described: `ode' is the solution to \eqref{eqn:simplethetaode} subject to \eqref{eqn:hr_nd} and `enthalpy' is the conserved finite volume method for solving \eqref{eqn:consenergyHndreduced}. In the thermodynamically balanced case, \eqref{eqn:hr_nd} can be integrated exactly and the fringe thickness can be found with a root-finding algorithm, which we name `root' in figure \ref{fig:hNcomparison}. The last two methods `uega' and `shooting' are from \citet{Rem2008} and are run here using the parameters in table \ref{prms}. In the uniform external gradient approximation, i.e. uega, method \citet{Rem2008} maps the fringe with imposed external heat fluxes to a Stefan problem domain whereas the `shooting' method searches for a consistent temperature at the base of the lowest ice lens. As expected, all 5 of these solution methods give the same result. The fringe thicknesses are much lower in the temperate melting case because the liquid pressure distribution in the fringe needed to expel meltwater supports a larger portion of the overburden and the fringe melts as the ice infiltrates into the sediments.

\begin{figure}
    \includegraphics[width=0.5\linewidth]{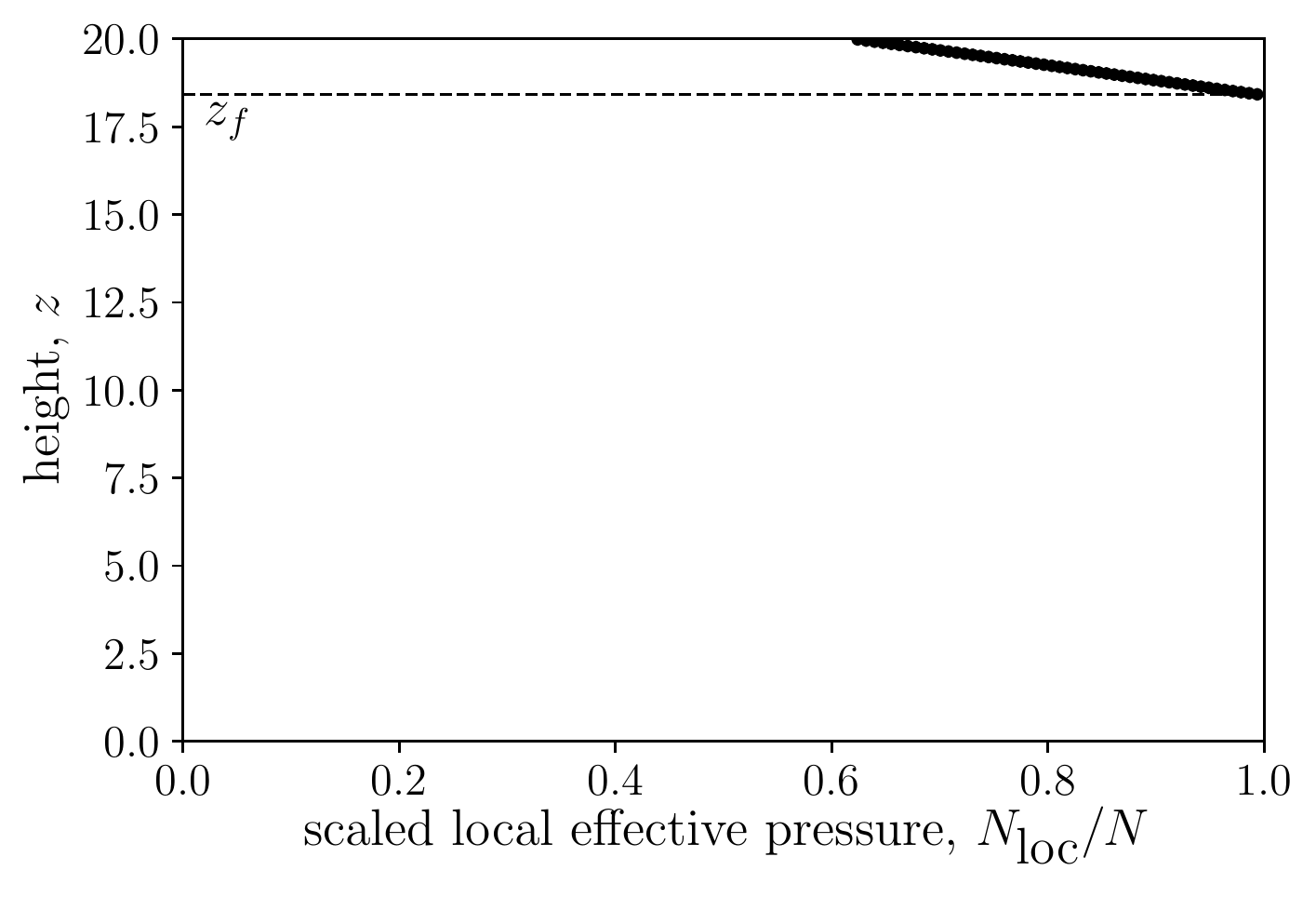}
    \includegraphics[width=0.5\linewidth]{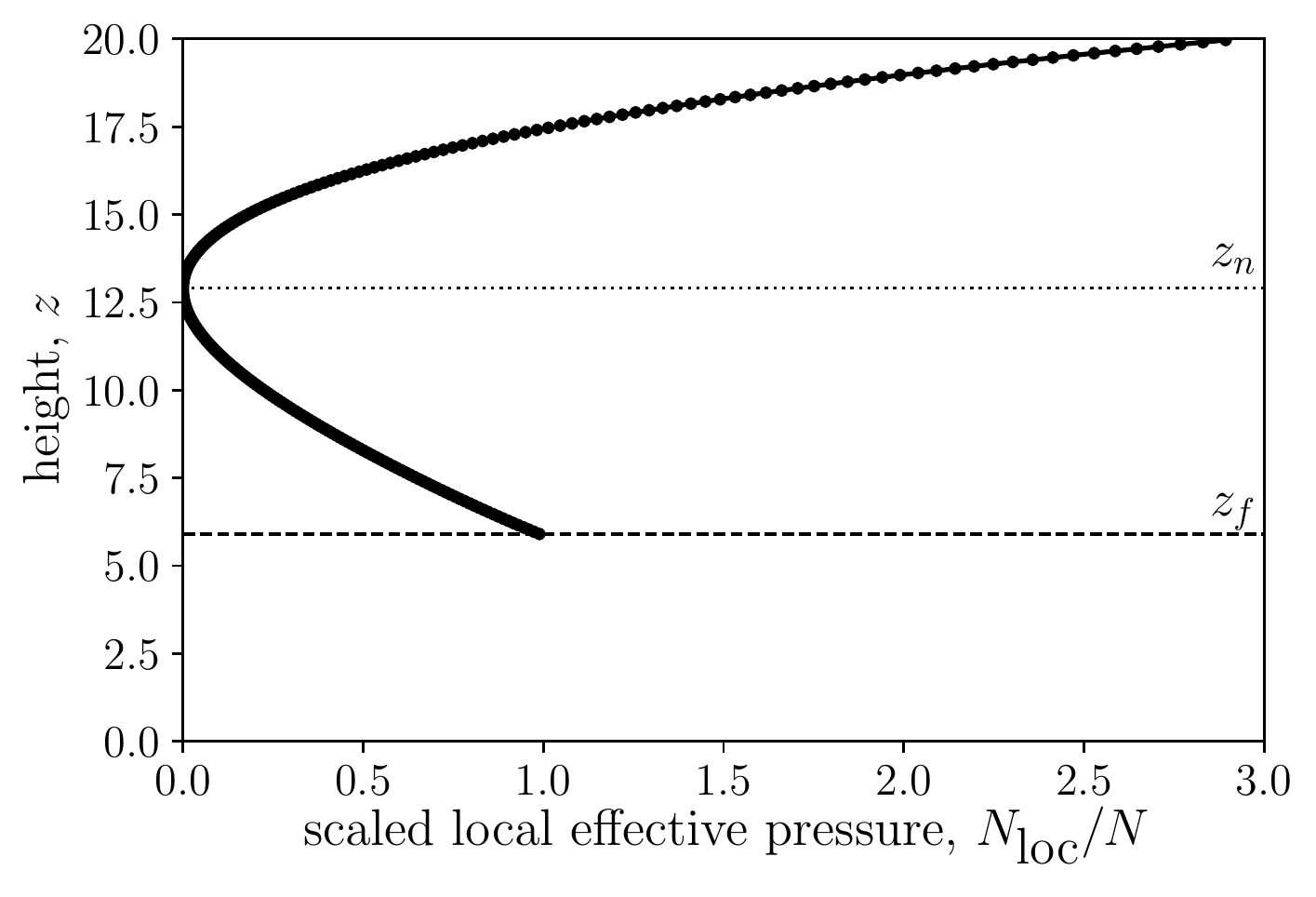}
    \caption{Local effective pressure $N_{\textrm{loc}}(z)$ with height $z$ scaled with the effective pressure $N$ at the base of the fringe for: (left) a melting steady state ($V=-0.01$ and $N=2.9$) where $N_{\textrm{loc}}>0$ throughout the fringe and (right) a transient freezing simulation ($V=0.20$ and $N=2.9$) where the local effective pressure goes to zero $N_{\textrm{loc}}(z_n)=0$, at point in the fringe $z_n$, initiating a new ice lens.}
    \label{fig:localeffp}
\end{figure}

\subsection{Ice lens initiation}
In the calculation of the steady state fringe thicknesses, we focused on melting ($V<0$) and balanced ($V=0$) thermodynamics. Although steady states do exist for relatively small freezing rates and relatively small effective pressures \citep{Mey2018c}, transient behaviour such as ice lens nucleation can occur when there is net freezing ($V>0$). In figure \ref{fig:localeffp}, we show the local effective pressure $N_{\textrm{loc}}$ as a function of depth $z$ for melting (left) and freezing (right). Here the top of the domain is the lowest ice lens $z_\ell = 20$ and $N_{\textrm{loc}}$ is evaluated in the fringe region above $z_f \approx 18$ (left) and $z_f \approx 6$ (right). In the steady melting case, $N_{\textrm{loc}}$ monotonically increases from the ice lens to the effective pressure at the bottom of the fringe $N$. The low pressure at the base of the ice lens draws in water as the ice lens infiltrates into the sediments through regelation \citep{Gil1980,Fow1994,Rem2019}. In the transient freezing case, water is drawn into the fringe and freezes onto the base of the lowest ice lens. The ice saturation increases, which lowers the permeability and requires a larger pressure difference to continue freezing. At some point in time, the local effective pressure reaches zero $N_{\textrm{loc}}(z_n)=0$ and a new ice lens forms at $z_n$. 

We determine the time when a new ice lens forms using the `events' functionality built into `solve\_ivp', which flags the location and time of the new ice lens as an event and stops the integration. We then shift our domain up so that the new ice lens is at the top and pad the bottom of the domain with water-saturated sediments following the same incoming heat flux. Then, we restart the integration until the next ice lens forms. Written inside a loop, we generate sequences of ice lenses with an interlens time $t_\ell$. 

\begin{figure}
\includegraphics[width=0.5\linewidth]{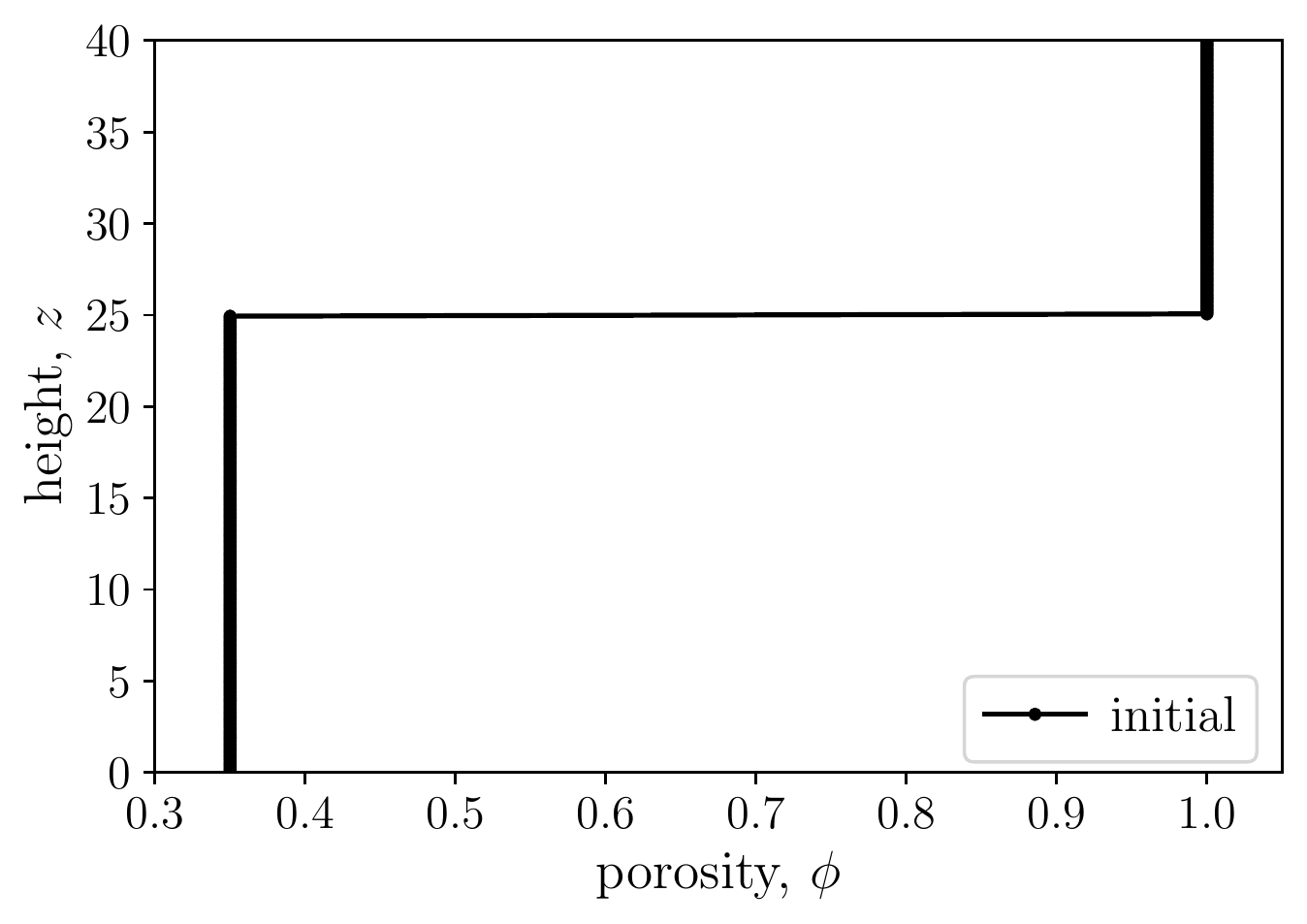}
\includegraphics[width=0.5\linewidth]{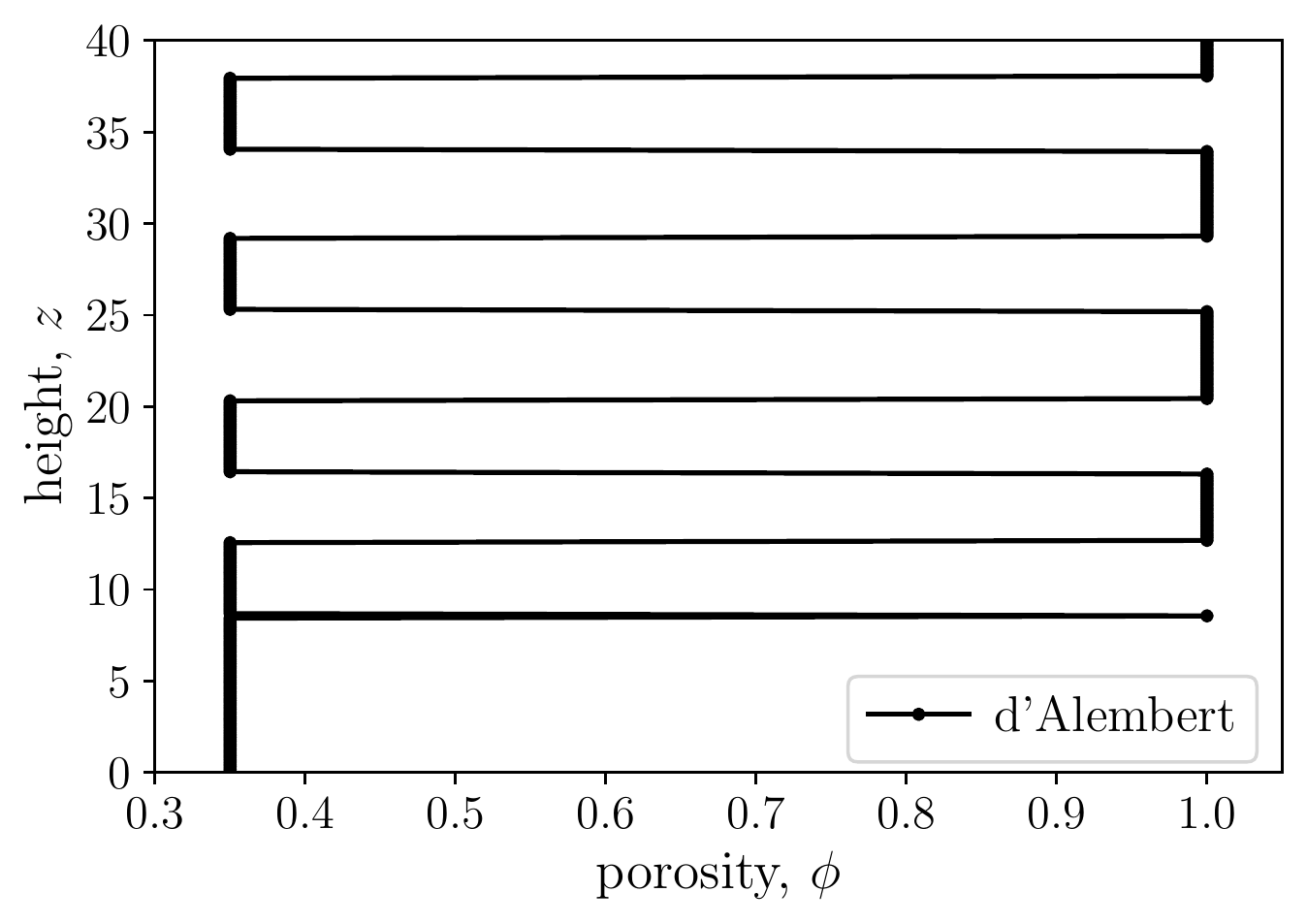}
\caption{Porosity structure of a freezing sediment pack: (left) initial porosity profile with $z_\ell=25$. The region above $z_\ell$ is the lowest ice lens and the region below is fringe as well as water-saturated sediments. (right) \emph{a posteriori} d'Alembert advection solutions for porosity after nucleation and growth of 4 periodic ice lenses. The nondimensional parameters $V=0.5$ and $N=1.5$ and the nondimensional interlens time is $t_{\ell}=9.6$.}
\label{fig:dAlembert}
\end{figure}

Using the constant heave rate $V$ and interlens time $t_\ell$, we can reconstruct the porosity structure \emph{a posteriori}. We treat the porosity as a constant $\phi$ in the fringe and water-saturated sediment. In the ice lenses, the porosity is also constant with $\phi=1$. In one vertical dimension, mass conservation for sediments from equation \eqref{sedcons} is
\begin{equation}
\pd{\left( 1-\phi\right)}{t} + \pd{\left[V_s \left( 1-\phi\right)\right]}{z} = 0,
\label{sedadv}
\end{equation}
for a constant sediment density $\rho_s$. If we say that the sediment advection $V_s$ is given by the heave according to
\begin{equation}V_s = 
\begin{cases}
0, & \mbox{below lowest lens} \\
V, & \mbox{above lowest lens}
\end{cases},
\label{vsdefine}
\end{equation}
we find that 
\begin{equation}
\pd{\phi}{t} + V \pd{\phi}{z} = 0,
\label{eqn:advphi}
\end{equation}
above $z_f$ since $V$ is constant in this region. This model assures that the ice lenses and interstitial fringe advect vertically as one unit of rigid ice \citep{O'Ne1985}. Importantly, this treatment assumes that there is no volume expansion upon freezing and the water freezes in place, requiring that ice and water have the same density. This is true to leading order in the scaled and simplified model we describe in \S \ref{reduction}. We solve equation \eqref{eqn:advphi} equation analytically using the d'Alembert form, i.e. 
\begin{equation}
\phi = f\left( z - V t\right),
\end{equation}
which we augment with the location of each new ice, thereby increase the region of applicability for equation \eqref{eqn:advphi} and use the fact that $\phi = 1$ at $z=z_f$. In figure \ref{fig:dAlembert}, we show the evolution of the porosity with time as 5 new ice lenses sequentially nucleate and grow. With these parameters, the lenses form periodically with equal spacing and interlens times. 

\begin{figure}
\centering
    \includegraphics[width=0.8\linewidth]{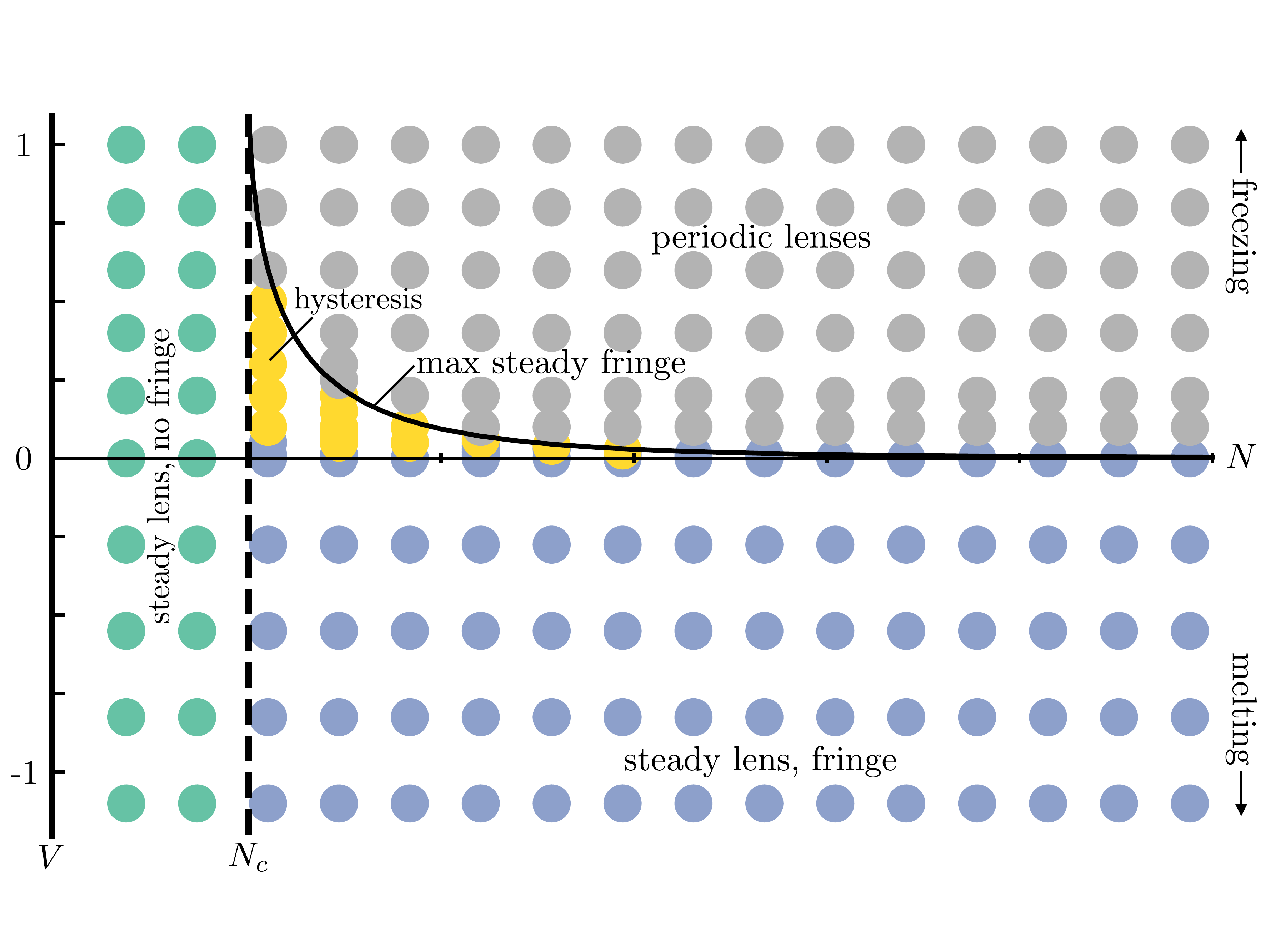}
    \caption{Regime diagram showing the system behaviour as a function of the heave rate $V$ and effective pressure $N$. The maximum steady fringe thickness is shown as a solid black line and theory is described in the appendix.}
    \label{fig:regime}
\end{figure}

As we have seen, the two primary control parameters for the system are the heave rate $V$ and the effective pressure $N$. So far we have shown steady states for balanced ($V=0$) and melting ($V<0$) conditions as well as ice lens nucleation for freezing conditions ($V>0$). In figure \ref{fig:regime}, we show a regime diagram for the system behaviour as a function of $V$ and $N$.  Each point on the figure is a simulation and the colour denotes the grouping. For $N<N_c$, no fringe forms and the lens either melts or grows, depending on the sign of $V$. When $N>N_c$, ice infiltrates into the sediment forming a fringe. In melting or balanced cases where $V\leq 0$, a steady state fringe thickness emerges (e.g. figures \ref{fig:enthalpyodecomparison} and \ref{fig:hNcomparison}). For positive heave rates $V>0$, steady states for relatively small effective pressures give way to periodic lenses as the heave rate increases. At the boundary between the steady and periodic regimes, there is a zone of hysteresis, where depending on the initial conditions the system either relaxes to a steady state or periodically generates new ice lenses. This same hysteresis was observed by \citet{Rem2004}. Although we observe some small variability in the interlens times, we do not see anything reminiscent of the chaotic regime found by \citet{And2014}. Based on figure \ref{fig:regime}, if we have estimates for $V$ and $N$ in a geophysical context such as below a glacier, we can predict the system behaviour such as whether ice lenses will form. 

\section{Conclusions\label{sec:conclusions}}
In this paper, we derived the thermodynamical and fluid mechanical equations governing frost heave in a geophysical context. We systematically reduced the equations using the fact that ice and water have similar densities (i.e. small $\delta$) as well as the large Stefan number $\mathtt{St}$. We solved the reduced set of equations using an enthalpy method where the interstitial ice saturation acts like a liquidus condition in the frozen fringe and conservation of enthalpy allows us to determine the fringe interface implicitly. For melting and balanced thermodynamics, we compared our enthalpy method to a steady state cast as an ordinary differential equation for temperature and found excellent agreement. In freezing cases, we found that the local effective pressure can go to zero within the fringe and nucleate a new ice lens. We accommodate this process in our enthalpy model using an `events' function that stops the integration when a new ice lens forms and restarts the integration with a domain shifted below the new ice lens. Based on our solutions for the time between lenses, we can reconstruct the porosity profile showing the sequence of ice lenses. Finally, we compiled a regime diagram of our simulation results, showing the onset of periodic lensing and behaviour including hysteresis. These results will inform the regime of geophysical systems and future investigations into the role of compaction as well as comparisons with laboratory experiments. 

\section{Acknowledgements}
We thank Andrew Fowler and Ian Hewitt for insightful conversations. We acknowledge support from NSF--2012958 (CRM and AWR) and NSF--1603907 (AWR) as well as ERDC/CRREL--W913E519C0008 (CRM). AWR and CS are grateful to have participated in the 2006 Geophysical Fluid Dynamics summer school at the Woods Hole Oceanographic Institution where part of this collaboration began. 

\appendix
\section{Steady heave near thermal balance}
For an effective pressure $N$ greater than the infiltration threshold $N_c$, a steady fringe will form for melting ($V<0$), balanced ($V=0$), and weakly freezing ($0<V<V_{\textrm{max}}(N)$) thermodynamics, where $V_{\textrm{max}}$ depends on the effective pressure $N$ (viz. figures \ref{fig:hNcomparison} and \ref{fig:regime}). Alternatively, the maximum freezing heave rate can be thought of in the reverse: for a given heave rate ($V>0$), there is a maximum effective pressure $N_{\textrm{max}}(V)$, which is the largest load that can be supported by a steady fringe. For larger effective pressures, periodic lenses form. Here we will show how to calculate $V_{\textrm{max}}(N)$ and $N_{\textrm{max}}(V)$.

We start by rearranging \eqref{eqn:hr_nd} for $N$, which is 
\begin{equation}
N  =1 + \mathtt{Gr}\left(\nu-1\right) (1-\phi) \left(z_{\ell}- z_f\right) 
+ \int_{z_f}^{z_{\ell}}{\left( 1 - \phi S\right)\pd{\theta}{z'}dz'} - V\int_{z_f}^{z_{\ell}}{\frac{(1-\phi S)^2}{k}dz'}.
\label{eqn:forbalappendix}
\end{equation}
To find the maximum effective pressure, we treat $z_f$ as fixed and set the derivative of $N$ with respect to $z_\ell$ equal to zero, i.e.
\begin{equation}
\pd{N}{z_\ell}  =\mathtt{Gr}\left(\nu-1\right) (1-\phi)
+ \left.\left( 1 - \phi S\right)\pd{\theta}{z}\right|_{z_\ell} - V\left.\frac{(1-\phi S)^2}{k}\right|_{z_\ell} = 0.
\end{equation}
Treating the thermal conductivity variation as negligible, we insert the heat flux from equation \eqref{eqn:simplethetaode} to find
\begin{equation}
\mathtt{Gr}\left(\nu-1\right) (1-\phi)
+ \left[ 1 - \phi S(\theta_\ell^*)\right] \left[1 + \mathtt{Pe}V\phi S(\theta_\ell^*)\right] - V\frac{\left[1-\phi S(\theta_\ell^*)\right]^2}{k(\theta_\ell^*)} = 0,
\end{equation}
which can be solved using a root-finding algorithm for the lens temperature $\theta_\ell^*(V)$, i.e. the lowest lens temperature that can be supported for a given heave rate $V$. We now change the integration variable from $z$ to $\theta$ in the force balance equation \eqref{eqn:forbalappendix} using equation \eqref{eqn:simplethetaode} and use the minimum lens temperature $\theta_{\ell}^*$ in the limits of integration, i.e.
\begin{eqnarray}
N  =1 + \int_{0}^{\theta_\ell^*(V)}{\frac{\mathtt{Gr}\left(\nu-1\right) (1-\phi)}{1 + \mathtt{Pe}V\phi S(\theta)}d\theta}  
+ \int_{0}^{\theta_\ell^*(V)}{\left[ 1 - \phi S(\theta)\right]d\theta} \hspace{3.7cm}\nonumber \\ - V\int_{0}^{\theta_\ell^*(V)}{\frac{[1-\phi S(\theta)]^2}{k(\theta) \left[1 + \mathtt{Pe}V\phi S(\theta)\right]}d\theta}.
\label{eqn:forbalchangevar}
\end{eqnarray}
For a given value of $N$, we use another root-finding algorithm to find the heave rate $V$ that satisfies equation \eqref{eqn:forbalchangevar}, which is the maximum heave rate with a steady fringe. Repeated application of this root-finding algorithm results in the black curve $V_{\textrm{max}}(N)$ shown on figure \ref{fig:regime} and partitions the periodic lens regime from the hysteresis/steady lens regime. 

\bibliographystyle{plainnat}
\bibliography{Glib}
\end{document}